\documentclass{article}

\usepackage[utf8]{inputenc} 
\usepackage[T1]{fontenc}    
\usepackage{hyperref}       
\usepackage{url}            
\usepackage{booktabs}       
\usepackage{amsfonts}       
\usepackage{nicefrac}       
\usepackage{microtype}      
\usepackage{graphicx}
\usepackage{natbib}
\usepackage{doi}
\usepackage{amsthm}
\usepackage{amsmath}
\usepackage{amssymb}
\newtheorem{theorem}{Theorem}
\usepackage[ruled,linesnumbered]{algorithm2e}
\usepackage{boldline} 
\usepackage{float}

\usepackage{appendix}
\usepackage{authblk}
\usepackage[a4paper, total={6in, 9in}]{geometry}

\usepackage[lofdepth,lotdepth,caption=false]{subfig}
\usepackage{fancyhdr}
\usepackage{xspace}
\usepackage{braket}
\usepackage{color}
\usepackage{setspace}
\usepackage{multirow} 
\usepackage[framemethod=tikz]{mdframed}
\usepackage{ctable}

\allowdisplaybreaks

\providecommand{\keywords}[1]
{
  \textbf{Keywords:} #1
}

\title{A Multiprocess State Space Model with Feedback and Switching for Patterns of Clinical Measurements Associated with COVID-19}

\date{} 					

\author[1]{\large Xiaoran Ma}
\author[2]{Wensheng Guo}
\author[3,4]{Peter Kotanko}
\author[1]{Yuedong Wang\thanks{Corresponding author: yuedong@pstat.ucsb.edu}}
\affil[1]{ \normalsize Department of Statistics and Applied Probability, University of California, Santa Barbara}
\affil[2]{Department of Biostatistics, Epidemiology and Informatics, University of Pennsylvania}
\affil[3]{Renal Research Institute}
\affil[4]{Icahn School of Medicine at Mount Sinai}

\begin{document}
\maketitle

\begin{abstract}
Clinical measurements, such as body temperature, are often collected over time to monitor an individual's underlying health condition. These measurements exhibit complex temporal dynamics, necessitating sophisticated statistical models to capture patterns and detect deviations. 
We propose a novel multiprocess state space model with feedback and switching mechanisms to analyze the dynamics of clinical measurements. This model captures the evolution of time series through distinct latent processes and incorporates feedback effects in the transition probabilities between latent processes. We develop estimation methods using the EM algorithm, integrated with multiprocess Kalman filtering and multiprocess fixed-interval smoothing. Simulation study shows that the algorithm is efficient and performs well. We apply the proposed model to body temperature measurements from COVID-19-infected hemodialysis patients to examine temporal dynamics and estimate infection and recovery probabilities.
\end{abstract}


\section{Introduction}
The COVID-19 pandemic has prompted extensive research into statistical methods for analyzing infection and recovery dynamics. 
Extensive existing literature has demonstrated the success of interpretable models in various prediction tasks. These models provide valuable insights into how diseases progress and help identify critical factors influencing recovery \citep{ferrari_machine_2020,Rustam_2020,ikemura_using_2021,song_predicting_2022,duan_predicting_2023,xiao_predicting_2024}. However, their application often faces challenges in capturing individual variability and adapting to real-time changes in patient conditions.
A critical aspect of this research is the analysis of trajectories of clinical measurements at the individual level. Although existing studies have identified significant changes in individual measurements, many models fail to accurately detect shifts in the underlying health condition \citep{chaudhuri_trajectories_2022,ma_nonparametric_2024}. Identifying and predicting these shifts is crucial for diagnosis and prognosis, assessing disease severity, and implementing preventive measures
 \citep{berzuini_value_2020,yu_novel_2020,fazal_c-reactive_2021,moisa_dynamic_2021}.

State space methods and regime-switching models have gained popularity in the analysis of complex time series data and have demonstrated potential in examining individual health records \citep{guo_signal_1999,guo_structural_2000,liu_modeling_2014,samdin_unified_2017,noman_markov-switching_2020}. Numerous studies have utilized these methods to investigate intervention effects \citep{kobayashi_predicting_2020}, forecast outcomes \citep{petrica_regime_2022,odea_semi-parametric_2022,noh_inference_2023}, track disease transmission \citep{zhou_semiparametric_2020,deo_new_2021,keller_tracking_2022}, and explore other socioeconomic factors related to COVID-19 at the population level \citep{shah_regime_2021}. However, methods for detecting change points at the individual level remain limited. In this paper, we propose a state space model with feedback and switching and apply it to study body temperature profiles of COVID-19-infected hemodialysis patients. The proposed model is an extension of the multiprocess state space model with Markov switching in \cite{kim_dynamic_1994} with an additional feedback mechanism. It is important to note that the inclusion of feedback makes the switching process non-Markovian, complicating the estimation. 

We focus on the transition of clinical measurements from one status to another in response to an event such as COVID-19 infection. Our model and method extend the existing literature in several ways. First, our model focuses on the feedback effect of transition probability rather than the feedback effect of an event process as in \cite{guo_structural_2000}. By characterizing changes in a subject's status, the transition probability serves as a tool for detecting change points and enabling online prediction. Second, we extend the multiprocess Kalman filtering and multiprocess fixed-interval smoothing procedures, including a detailed discussion on handling missing data, a common issue in longitudinal studies where some observations may be unavailable. Third, we derive efficient approximations to achieve a linear time complexity.



\begin{figure}[H]
    \centering
    \includegraphics[width=\textwidth]{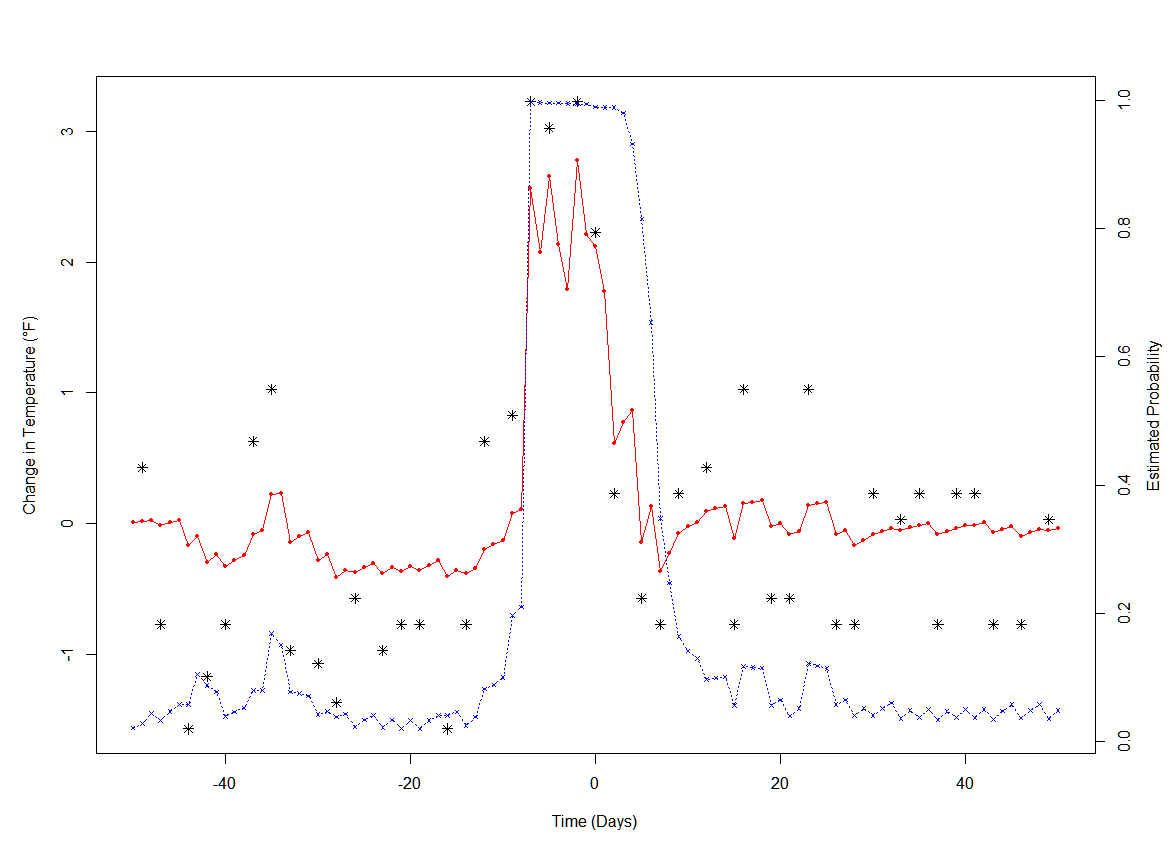}
    \caption{Change in temperature from the baseline and estimates of a COVID-19-infected hemodialysis patient. Time 0 corresponds to the date when COVID-19 infection was confirmed based on a positive RT-PCR test. The black star points are the observed change in temperatures. The red solid line with dots is the filtering estimate of the underlying body temperature changes. The blue dashed line with crosses is the filtering estimate of probabilities for having a fever.}
    \label{MSSFS_fig_pre}
\end{figure}

As an illustration of the application of the proposed method, for a COVID-19-infected hemodialysis patient, Figure \ref{MSSFS_fig_pre} shows the change in body temperature from the baseline (exact definition see Section \ref{MSSFS_RDA}) and estimates of the underlying changes and the probabilities of having a fever. 


The rest of the paper is organized as follows. We introduce the Multiprocess State Space Model with Feedback and Switching (MSSFS) and estimation procedure in Sections \ref{MSSFS_model} and \ref{MSSFS_estimation}. The real data analysis and simulation results are reported in Sections \ref{MSSFS_RDA} and \ref{MSSFS_sim}. Conclusions are presented in Section \ref{conclusion}.

\section{Model Specification} \label{MSSFS_model}
Denote the observation vector and the latent state vector for the $i$-th subject at time $t$ as $\boldsymbol{y}_{i,t} \in \mathbb{R}^{p}$ and $\boldsymbol{\theta}_{i,t} \in \mathbb{R}^{q}$, where $i = 1, 2, \ldots, m$ and $t = 1, \ldots, n$. 
Denote the stacked vector of the response and latent state as $\boldsymbol{y}_{i,1:t} = (\boldsymbol{y}_{i,1}^T, \ldots, \boldsymbol{y}_{i,t}^T)^T$ and $\boldsymbol{\theta}_{i,1:t} = (\boldsymbol{\theta}_{i,1}^T, \ldots, \boldsymbol{\theta}_{i,t}^T)^T$, respectively. 
To better represent real-world scenarios, we allow certain elements or even the entire vector $\boldsymbol{y}_{i,t}$ to be missing. 

The Multiprocess State Space model with Feedback and Switching (MSSFS) extends the regular state space model by allowing the system equation to vary according to the underlying status of the subject at time $t$. For simplicity, we consider two statuses, and our model and methods can be extended to situations with more than two statuses. Denote $I_{i,t} \in \{0,1\}$ as the status for the $i$-th subject at time $t$. 
An MSSFS model assumes that
\begin{align}
&\text{Observation Equation:} &\boldsymbol{y}_{i,t} &=  \mathbf{F}_{i,t} \boldsymbol{\theta}_{i,t}+\boldsymbol{v}_{i,t} , \label{OE}  \\
&\text{System Equations:} &\boldsymbol{\theta}_{i,t} &=  \boldsymbol{\gamma}_{i,t,0}  + \mathbf{G}_{i,t,0} \boldsymbol{\theta}_{i,t-1}+\boldsymbol{w}_{i,t,0}   \quad \text{if }I_{i,t}=0
\label{SE1},\\
&  &\boldsymbol{\theta}_{i,t} &=  \boldsymbol{\gamma}_{i,t,1} + \mathbf{G}_{i,t,1}  \boldsymbol{\theta}_{i,t-1}+\boldsymbol{w}_{i,t,1}  \quad \text{if }I_{i,t}=1
\label{SE2},\\
&\text{Switch Equations:} & \pi_{i,t,01} &= \frac{\exp \{\alpha_0 + \boldsymbol{x}_i^T\boldsymbol{\beta}_0 + z_{i,t,0}(\boldsymbol{\theta}_{i,1:(t-1)})  \}}{1+\exp\{\alpha_0 + \boldsymbol{x}_i^T\boldsymbol{\beta}_0 + z_{i,t,0}(\boldsymbol{\theta}_{i,1:(t-1)})  \}} \label{SE3}, \\
&                        & \pi_{i,t,11} &= \frac{\exp \{\alpha_1 + \boldsymbol{x}_i^T\boldsymbol{\beta}_1 + z_{i,t,1}(\boldsymbol{\theta}_{i,1:(t-1)})  \}}{1+\exp\{\alpha_1 + \boldsymbol{x}_i^T\boldsymbol{\beta}_1 + z_{i,t,1}(\boldsymbol{\theta}_{i,1:(t-1)}) \}} ,\label{SE4} 
\end{align}
where $\pi_{i,t,op} = \operatorname{Pr}[I_{i,t}=p|I_{i,t-1}=o]$ with $o,p \in \{0,1\}$ denote the switching probabilities. 
The observation equation \eqref{OE} represents the relation between the response and the latent states $\boldsymbol{\theta}_{i,t}$, where $\mathbf{F}_{i,t} \in \mathbb{R}^{p \times q}$ is a matrix and $\boldsymbol{v}_{i,t} \in \mathbb{R}^{p}$ is a vector of random observation errors. 
The system equations \eqref{SE1} and \eqref{SE2} characterize how the latent states evolve over time under different statuses, where $\boldsymbol{\gamma}_{i,t,k} \in \mathbb{R}^{q}$ are drifting vectors, $\mathbf{G}_{i,t,k} \in \mathbb{R}^{q \times q}$ are matrices, and $\boldsymbol{w}_{i,t,k} \in \mathbb{R}^{q}$ are vectors of random disturbances.
We assume that $\boldsymbol{v}_{i,t} \sim N\left(\boldsymbol{0}, \mathbf{V}_{i,t}\right)$ and $\boldsymbol{w}_{i,t,k} \sim N\left(0, \mathbf{W}_{i,t,k}\right)$, and they are mutually and serially independent and are independent across different subjects. 
We also assume they are independent of the underlying processes $\boldsymbol{\theta}_{i,t}$. 
The switch equations \eqref{SE3} and \eqref{SE4} model the transition probabilities using logistic regression models with covariates $\boldsymbol{x}$ and a function of latent states up to the previous time point, $z_{i,t,k}(\boldsymbol{\theta}_{i,1:(t-1)})$. 
We note that some elements in the covariates $\boldsymbol{x}$ may depend on time, which was not expressed explicitly for simplicity of notation. The time-varying effect,
$z_{i,t,k}(\boldsymbol{\theta}_{i,1:(t-1)})$, 
can be any function of the previous latent states that provides a flexible mechanism to model feedback through transition probabilities. 
 For example, one may consider a simple linear function $z_{i,t,k}(\boldsymbol{\theta}_{i,1:(t-1)}) =  \boldsymbol{\zeta}_k^T \boldsymbol{\theta}_{i,(t-L):(t-1)}$, where $L$ is a positive integer and $\boldsymbol{\zeta}_k  \in \mathbb{R}^{q}$ are the parameters to be estimated.


For initialization of the model, we assume a normal distribution for the initial distribution of the latent state, $\boldsymbol{\theta}_{i,0} \stackrel{iid}{\sim} N(\boldsymbol{\mu}_i,\mathbf{\Sigma}_i)$, where the mean vector $\boldsymbol{\mu}_i$ and covariance matrix $\mathbf{\Sigma}_i$ can be regarded as parameters to be estimated. Another way is to use diffuse initialization as in \cite{guo_structural_2000} and \cite{durbin_time_2012}. More details on the initialization for the real data applications will be discussed in Section \ref{MSSFS_RDA}.

\section{Model Estimation} \label{MSSFS_estimation}
\subsection{EM algorithm}

Since $z_{i,t,k}$ in the transition probability model is not observable, we apply the EM algorithm to estimate the parameters and latent states simultaneously. For each subject $i$, let $\boldsymbol{z}_{i,t} = ( z_{i,t,0}(\boldsymbol{\theta}_{i,1:(t-1)}), z_{i,t,1}(\boldsymbol{\theta}_{i,1:(t-1)}) )^T$ and $\boldsymbol{z}_{i,1:t} = (\boldsymbol{z}_{i,1}^T, \ldots, \boldsymbol{z}_{i,t}^T)^T$. We use $\boldsymbol{I}_{i,1:t} = (I_{i,1}, \ldots, I_{i,t})^T$ to denote the statuses for subject $i$ up to time $t$. The algorithm regards $\boldsymbol{z}_{i,1:n}$ as missing values and uses the EM algorithm as in \cite{guo_structural_2000} to estimate the parameters. We assume there are no missing values in covariates $\boldsymbol{x}_i$. We develop methods in Section \ref{MSSFS_estimation_MKF} to deal with missing values in $\boldsymbol{y}_{i,1:n}$.  Let $\boldsymbol{y} = (\boldsymbol{y}_{1,1:n}^T, \ldots, \boldsymbol{y}_{m,1:n}^T)^T$ and $\boldsymbol{z} = ( \boldsymbol{z}_{1,1:n}^T,\ldots,\boldsymbol{z}_{m,1:n}^T )^T$.
 Denote $\boldsymbol{\Theta}$ as all parameters to be estimated. The EM algorithm is described in Algorithm \ref{MSSFS_algo}. The details of Multiprocess Kalman Filtering (MKF) and Multiprocess Fixed Interval Smoothing (MFIS) in the algorithm are discussed in Sections \ref{MSSFS_estimation_MKF} and \ref{MSSFS_estimation_MFIS} respectively.

\IncMargin{1em}
\begin{algorithm} 
\label{MSSFS_algo}
\SetAlgoLined
Derive initial estimates of $\hat{\boldsymbol{\Theta}}^0$ and $\hat{\boldsymbol{\theta}}^0$ by using MKF and MFIS without the term involving $z_{i,t,k}(\cdot)$ in equations (\ref{SE3}) and (\ref{SE4}). Then plug $\hat{\boldsymbol{\theta}}^0$ into the function to get $\hat{\boldsymbol{z}}^0$\;
\While{$(\{d_{EM}^{\tau} > D_{EM}\} \wedge \{\tau \leq \mathcal{N}_{max}\})$}{
\textbf{E-step:} \\
(a) With the previous estimate of 
$\hat{\boldsymbol{z}}^{\tau-1}$, compute $Q$ in equation \eqref{Q_approx} using MKF\; 
\textbf{M-step:} \\
(b) Maximize $Q$ to get an update of $\boldsymbol{\Theta}$, $\hat{\boldsymbol{\Theta}}^{\tau}$\; 
(c) With $\hat{\boldsymbol{\Theta}}^{\tau}$, compute $\hat{\boldsymbol{\theta}}^{\tau}$ using MFIS\; 
\textbf{Plug-in Estimate:} \\
(d) With $\hat{\boldsymbol{\theta}}^{\tau}$, compute $\hat{\boldsymbol{z}}^{\tau}$ using plug-in estimate.
}
\caption{EM algorithm for fitting the MSSFS model}
\end{algorithm}
\DecMargin{1em}

\begin{figure}[H]
    \centering    
    \includegraphics[width=\textwidth]{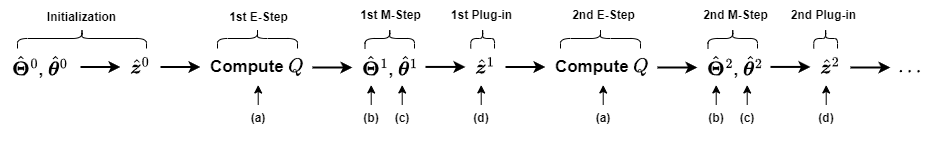}
    \caption{Flow chart of the EM algorithm.}
    \label{MSSFS_EM}
\end{figure}

Figure \ref{MSSFS_EM} provides the flow chart of the EM algorithm. We start the algorithm by fitting a model without feedback $\boldsymbol{z}$ to derive initial estimates  $\hat{\boldsymbol{\Theta}}^0$ and $\hat{\boldsymbol{\theta}}^0$, where the superscript denotes the iteration number. We then plug $\hat{\boldsymbol{\theta}}^0$ into $z$ function to get $\hat{\boldsymbol{z}}^0$. With $\boldsymbol{z}$ fixed at $\hat{\boldsymbol{z}}^0$, we perform the following steps in the first iteration: (a) compute approximated conditional expectation of the log-likelihood as in equation \eqref{Q_approx}; (b) maximize $Q$ to get an update of parameters $\hat{\boldsymbol{\Theta}}^1$; (c) with the updated parameters, apply MFIS to update the states, $\hat{\boldsymbol{\theta}}^1$; and (d) plug $\hat{\boldsymbol{\theta}}^1$ into functions to update $z$ values, 
$\hat{\boldsymbol{z}}^1$. The algorithm alternates among these steps until convergence.

The E-step involves the conditional expectation of the log-likelihood that does not have a closed form. We derive the following approximation to ease the computation. Let $\hat{\boldsymbol{\Theta}}^{\tau-1}$ denote all the parameters estimated at the previous iteration. We approximate the conditional expectation of the log-likelihood as follows:
\begin{align}
Q\left(\boldsymbol{\Theta} \mid \hat{\boldsymbol{\Theta}}^{\tau-1}, \boldsymbol{y}\right)
=& \int l(\boldsymbol{\Theta} \mid \boldsymbol{y}, \boldsymbol{z}) p\left(\boldsymbol{z} \mid \hat{\boldsymbol{\Theta}}^{\tau-1}, \boldsymbol{y}\right) d \boldsymbol{z} \nonumber\\
\approx & \int\left[l(\boldsymbol{\Theta} \mid \boldsymbol{y}, \hat{\boldsymbol{z}}^{\tau-1})+\left.\frac{\partial l}{\partial \boldsymbol{z}}\right|_{\hat{\boldsymbol{z}}^{\tau-1}}(\boldsymbol{z}-\hat{\boldsymbol{z}}^{\tau-1})\right] \times p\left(\boldsymbol{z} \mid \hat{\boldsymbol{\Theta}}^{\tau-1}, \boldsymbol{y}\right) d \boldsymbol{z} \nonumber\\
\approx& l(\boldsymbol{\Theta} \mid \boldsymbol{y}, \hat{\boldsymbol{z}}^{\tau-1}) \nonumber\\
=& \sum_{i=1}^{m} l(\boldsymbol{\Theta} \mid \boldsymbol{y}_{i,1:n}, \hat{\boldsymbol{z}}_{i,1:n}^{\tau-1}), \label{Q_approx}
\end{align}
where $l(\boldsymbol{\Theta} \mid \boldsymbol{y}_{i,1:n}, \hat{\boldsymbol{z}}_{i,1:n}^{\tau-1})$ is the logarithm of the joint density of $\boldsymbol{y}_{i,1:n}$ and $\boldsymbol{z}_{i,1:n}$ with $\hat{\boldsymbol{z}}_{i,1:n}^{\tau-1}$ plugged in as the observation. 
The first approximation is based on the first-order Taylor expansion.  
The second approximation holds because $\left.\frac{\partial l}{\partial \boldsymbol{z}}\right|_{\hat{\boldsymbol{z}}^{\tau-1}} \approx \boldsymbol{0}$ when $\boldsymbol{\Theta}$ is in the neighborhood of $\hat{\boldsymbol{\Theta}}^{\tau-1}$.
To see this, we use the fact that 
$\hat{\boldsymbol{\theta}}^{\tau-1}$ maximizes $\log p(\boldsymbol{\theta} \mid \boldsymbol{y}, \hat{\boldsymbol{\Theta}}^{\tau-1})$  
and consequently, $\hat{\boldsymbol{z}}^{\tau-1}$ 
maximizes 
$\log p(\boldsymbol{z} \mid \boldsymbol{y}, \hat{\boldsymbol{\Theta}}^{\tau-1})$ since $\boldsymbol{z}$ is a function of $\boldsymbol{\theta}$. 
We further use the fact that
$l(\boldsymbol{\Theta} \mid \boldsymbol{y}, \boldsymbol{z}) = \log p(\boldsymbol{z}, \boldsymbol{y} \mid  \boldsymbol{\Theta}) = \log p(\boldsymbol{z} \mid \boldsymbol{y}, \boldsymbol{\Theta}) + \log p(\boldsymbol{y} \mid \boldsymbol{\Theta})$. Therefore,  $\left.\frac{\partial l}{\partial \boldsymbol{z}}\right|_{\hat{\boldsymbol{z}}^{\tau-1}} = \boldsymbol{0}$ when $\boldsymbol{\Theta}=\hat{\boldsymbol{\Theta}}^{\tau-1}$. The approximation follows based on the assumption that the log-likelihood is a smooth function of $\boldsymbol{\Theta}$.  
The last equality follows since we assume subjects are independent.
Even though the $Q$ function in the EM algorithm usually depends on the previous estimates $\hat{\boldsymbol{\Theta}}^{\tau-1}$, due to our approximation, the approximated $Q$ function only depends on $\hat{\boldsymbol{z}}^{\tau-1}$.
The M-step obtains $\boldsymbol{\Theta}$ that maximizes $Q\left(\boldsymbol{\Theta} \mid \hat{\boldsymbol{\Theta}}^{\tau-1}, \boldsymbol{y}\right)$ using the above approximation.
Simulations in Section \ref{MSSFS_sim} show that even though the approximation requires $\boldsymbol{\Theta}$ to be close to $\hat{\boldsymbol{\Theta}}^{\tau}$ for all $\tau$, it is robust to the choices of parameter initial values. 

The stopping criteria are set as follows. 
Let $d_{EM}^{\tau} =  \frac{\|\hat{\boldsymbol{\Theta}}^{\tau}-\hat{\boldsymbol{\Theta}}^{\tau-1} \|^2}{\| \hat{\boldsymbol{\Theta}}^{\tau-1}  \|^2 + \kappa} $ be the relative change of the estimates at the $\tau$-th EM iteration, and $\kappa$ is a small positive number to prevent underflow and overflow. The EM iteration stops if $d_{EM}^{\tau} \leq D_{EM}$, a predefined threshold, or reaches the maximum number of iterations $\mathcal{N}_{max}$.

\subsection{Multiprocess Kalman Filtering (MKF)} \label{MSSFS_estimation_MKF}

In this section, we develop Multiprocess Kalman Filtering (MKF) used in step (a) of the EM algorithm. It extends the Kalman filter in \cite{kim_dynamic_1994} by including the feedback effects and covariates. The feedback effects are not observed, and their estimates from the previous iteration are used in the switch equations. \cite{Harrison1976} presented the issue of computation time for multiprocess state space models and proposed collapsing, an approximation method to ease the computation. Assuming we have $K$ different statuses at each time point $t$, then the resultant posterior distribution will consist of a weighted combination of $K$ distinct normal distributions. Passing them to time $t+1$ and applying again the Kalman filter algorithm for each possible state to each component of the time $t$ posterior will generate a distribution comprising $K^2$ distinct normal components. The number of possible states to consider grows exponentially as $t$ increases and becomes intractable after a few steps. As in \cite{kim_dynamic_1994}, we apply the collapsing method to derive an approximation of $l(\boldsymbol{\Theta} \mid \boldsymbol{y}_{i,1:n}, \hat{\boldsymbol{z}}_{i,1:n}^{\tau-1})$ involved in  equation \eqref{Q_approx} as follows:
 
\begin{align}
&l(\boldsymbol{\Theta} \mid \boldsymbol{y}_{i,1:n}, \hat{\boldsymbol{z}}_{i,1:n}^{\tau-1}) \nonumber\\
= &
\log \left[p(\boldsymbol{y}_{i,1:n} \mid  \hat{\boldsymbol{z}}_{i,1:n}^{\tau-1}, \boldsymbol{\Theta}) p(\hat{\boldsymbol{z}}_{i,1:n}^{\tau-1} \mid \boldsymbol{\Theta}) \right] \nonumber\\
\approx & 
\log p(\boldsymbol{y}_{i,1:n} \mid  \hat{\boldsymbol{z}}_{i,1:n}^{\tau-1}, \boldsymbol{\Theta}) + \text{constant} \nonumber\\
\propto &
\log \left[ p(\boldsymbol{y}_{i,1} \mid \hat{\boldsymbol{z}}_{i,1:n}^{\tau-1},\boldsymbol{\Theta})  p(\boldsymbol{y}_{i,2} \mid \boldsymbol{y}_{i,1}, \hat{\boldsymbol{z}}_{i,1:n}^{\tau-1},\boldsymbol{\Theta})  \ldots p(\boldsymbol{y}_{i,n} \mid \boldsymbol{y}_{i,1:n-1}, \hat{\boldsymbol{z}}_{i,1:n}^{\tau-1},\boldsymbol{\Theta}) \right] \nonumber\\
\approx & 
\log \left[\sum_{I_{i,1}} p(\boldsymbol{y}_{i,1} \mid I_{i,1}, \hat{\boldsymbol{z}}_{i,1:n}^{\tau-1}, \boldsymbol{\Theta}) \operatorname{Pr}(I_{i,1} \mid \hat{\boldsymbol{z}}_{i,1:n}^{\tau-1}, \boldsymbol{\Theta})\right] \nonumber\\
& + \log\left[\sum_{I_{i,2}} \sum_{I_{i,1}} p(\boldsymbol{y}_{i,2} \mid  \boldsymbol{y}_{i,1}, \hat{\boldsymbol{z}}_{i,1:n}^{\tau-1},I_{i,2}, I_{i,1}, \boldsymbol{\Theta}) \operatorname{Pr}(I_{i,2}, I_{i,1} \mid \boldsymbol{y}_{i,1}, \hat{\boldsymbol{z}}_{i,1:n}^{\tau-1}, \boldsymbol{\Theta}) \right] \nonumber\\
& +
\ldots \nonumber\\
& + \log \left[ \sum_{I_{i,n}} \sum_{I_{i,n-1}} p(\boldsymbol{y}_{i,n} \mid \boldsymbol{y}_{i,1:n-1}, \hat{\boldsymbol{z}}_{i,1:n}^{\tau-1}, I_{i,n}, I_{i,n-1}, \boldsymbol{\Theta}) \operatorname{Pr}(I_{i,n}, I_{i,n-1} \mid \boldsymbol{y}_{i,1:n-1}, \hat{\boldsymbol{z}}_{i,1:n}^{\tau-1}, \boldsymbol{\Theta}) \right], \label{MSSFS_likelihood_approx}
\end{align}
where the first approximation assumes that $p(\hat{\boldsymbol{z}}_{i,1:n}^{\tau-1} \mid \boldsymbol{\Theta})$ is approximately independent of $\boldsymbol{\Theta}$, which is supported by the empirical evidence from simulations. The second approximation approximates the conditional event of all previous status $\boldsymbol{I}_{i,1:t}$ by the status at the two proceeding time points $\boldsymbol{I}_{i,(t-1):t}$. To get the probability $\operatorname{Pr}(I_{i,t}, I_{i,t-1} \mid \boldsymbol{y}_{i,1:t-1}, \hat{\boldsymbol{z}}_{i,1:t}^{\tau-1}, \boldsymbol{\Theta})$ for $t \in \{1,\ldots,n\}$, a collapse step in MKF is used to calculate $\operatorname{Pr}(I_{i,t-1} \mid \boldsymbol{y}_{i,1:t-1}, \hat{\boldsymbol{z}}_{i,1:n}^{\tau-1}, \boldsymbol{\Theta})$. This is used by \cite{kim_dynamic_1994} and has achieved reasonably good accuracy with a great computational advantage. The mathematical details of collapsing can be found in Theorem \ref{MSSFS_MKF} and its proof.

Since all subjects share the same log-likelihood structure and are independent of each other, we simplify the derivation of MKF and MFIS below for a single subject and drop the subscript $i$. 


Let $\boldsymbol{\psi}_t = (\boldsymbol{y}_{1:t}, \hat{\boldsymbol{z}}_{1:n}^{\tau-1}, \boldsymbol{x})$ where $ \hat{\boldsymbol{z}}_{1:n}^{\tau-1} $ is the estimate from the previous iteration in the EM algorithm for any value $\tau \ge 1$. Both $ \hat{\boldsymbol{z}}_{1:n}^{\tau} $ and covariates $\boldsymbol{x}$ are known for all time $t$.  

Consider the latent process $\boldsymbol{\theta}_t$ conditional on $\boldsymbol{\psi}_t$ and the status of the subject at time $t$ and $t-1$. Denote the conditional expectation and variance as
$\boldsymbol{\theta}_{t \mid t-1}^{(o,p)}= \mathbb{E}[\boldsymbol{\theta}_{t}  \mid  \boldsymbol{\psi}_{t-1}, I_{t-1}=o, I_{t}=p ]$
and  
$\boldsymbol{P}_{t \mid t-1}^{(o,p)}
= \operatorname{Cov}[\boldsymbol{\theta}_{t}  \mid  \boldsymbol{\psi}_{t-1}, I_{t-1}=o, I_{t}=p]$ 
respectively, where $o, p \in \{0,1\}$. The following theorem describes the details of the MKF.

\begin{theorem} \label{MSSFS_MKF}
    With appropriate initializations, the Multiprocess Kalman Filtering with $I_t \in \{0,1\}$ consists of the following steps.      
    \bigbreak
    For $t=1,\ldots,n $ DO:
    \begin{enumerate}        
        \item Start with \label{start}\\
        $\boldsymbol{\theta}_{t-1 \mid t-1}^{(o)} = \mathbb{E}\left[\boldsymbol{\theta}_{t-1} \mid I_{t-1}=o, \boldsymbol{\psi}_{t-1}\right]$, \\
        $\boldsymbol{P}_{t-1 \mid t-1}^{(o)} = \operatorname{Cov}\left[\boldsymbol{\theta}_{t-1} \mid I_{t-1}=o, \boldsymbol{\psi}_{t-1}\right]$, \\
        $\operatorname{Pr}(I_{t-1}=o \mid \boldsymbol{\psi}_{t-1})$.

        \item Calculate one-step forward predictions \\
        $\boldsymbol{\theta}_{t \mid t-1}^{(o,p)} = \boldsymbol{\gamma}_{p,t} + \mathbf{G}_{p,t} \boldsymbol{\theta}_{t-1 \mid t-1}^{(o)}$, \\
        $
            \boldsymbol{P}_{t \mid t-1}^{(o,p)} = \mathbf{G}_{p,t}  \boldsymbol{P}_{t-1 \mid t-1}^{(o)}  \mathbf{G}_{p,t}^T + \mathbf{W}_{p,t}
        $, \\
        $
                \boldsymbol{y}_{t \mid t-1}^{(o,p)} 
                = \mathbf{F}_t \boldsymbol{\theta}_{t \mid t-1}^{(o,p)}
        $.

        \item Calculate one-step observation prediction error and prediction error variance \\
        $
            \boldsymbol{\eta}_{t \mid t-1}^{(o,p)} = \boldsymbol{y}_t - \boldsymbol{y}_{t \mid t-1}^{(o,p)} 
        $,   \\
        $
            \mathbf{H}_{t \mid t-1}^{(o,p)} = \mathbf{F}_{t} \boldsymbol{P}_{t \mid t-1}^{(o,p)} \mathbf{F}_{t}^T + \mathbf{V}_t
        $.

        \item Calculate the posterior of states and their variances \label{MKF_4}\\
        $
            \boldsymbol{\theta}_{t \mid t}^{(o,p)} 
            \approx \boldsymbol{\theta}_{t \mid t-1}^{(o,p)} + \boldsymbol{P}_{t \mid t-1}^{(o,p)} \mathbf{F}_{t}^T \left( \mathbf{F}_{t} \boldsymbol{P}_{t \mid t-1}^{(o,p)} \mathbf{F}_{t}^T + \mathbf{V}_t \right)^{-1} \boldsymbol{\eta}_{t \mid t-1}^{(o,p)}
        $, \\
        $
            \boldsymbol{P}_{t \mid t}^{(o,p)} 
            \approx \boldsymbol{P}_{t \mid t-1}^{(o,p)} - \boldsymbol{P}_{t \mid t-1}^{(o,p)} \mathbf{F}_{t}^T \left( \mathbf{F}_{t} \boldsymbol{P}_{t \mid t-1}^{(o,p)} \mathbf{F}_{t}^T + \mathbf{V}_t \right)^{-1} \mathbf{F}_{t} \boldsymbol{P}_{t \mid t-1}^{(o,p)}
        $.

        \item Update probabilities \label{like}\\
        $\operatorname{Pr}(I_{t-1}=o, I_{t}=p \mid \boldsymbol{\psi}_{t-1}) = \operatorname{Pr}(I_{t}=p \mid I_{t-1}=o,  \boldsymbol{\psi}_{t-1}) \operatorname{Pr}(I_{t-1}=o \mid \boldsymbol{\psi}_{t-1})$,
         \\
        $
            \operatorname{Pr}(I_{t-1}=o, I_{t}=p \mid \boldsymbol{\psi}_{t}) 
            = \\
            \frac{(2 \pi)^{-n/2} |\mathbf{H}_{t \mid t-1}^{(o,p)}|^{-1/2} \exp(-\frac{1}{2} {\boldsymbol{\eta}_{t|t-1}^{(o,p)}}^T (\mathbf{H}_{t \mid t-1}^{(o,p)})^{-1} \boldsymbol{\eta}_{t|t-1}^{(o,p)}) \cdot \operatorname{Pr}(I_{t-1}=o, I_{t}=p \mid \boldsymbol{\psi}_{t-1})}{\sum_{p=0}^{1} \sum_{o=0}^{1} (2 \pi)^{-n/2} |\mathbf{H}_{t \mid t-1}^{(o,p)}|^{-1/2} \exp(-\frac{1}{2} {\boldsymbol{\eta}_{t|t-1}^{(o,p)}}^T (\mathbf{H}_{t \mid t-1}^{(o,p)})^{-1} \boldsymbol{\eta}_{t|t-1}^{(o,p)}) \cdot \operatorname{Pr}(I_{t-1}=o, I_{t}=p \mid \boldsymbol{\psi}_{t-1})}
        $, 
        \smallbreak
        \noindent
        $
            \operatorname{Pr}(I_{t}=p \mid \boldsymbol{\psi}_{t}) = \sum_{o=0}^{1} \operatorname{Pr}(I_{t-1}=o, I_{t}=p \mid \boldsymbol{\psi}_{t}) 
        $.

        \item Collapse statuses at time $t-1$ \label{collapse} \\
        $    
    \boldsymbol{\theta}_{t \mid t}^{(p)}
            =
            \sum_{o=0}^{1} \operatorname{Pr}(I_{t-1}=o \mid I_{t}=p, \boldsymbol{\psi}_{t}) \boldsymbol{\theta}_{t \mid t}^{(o,p)} = \frac{\sum_{o=0}^{1} \operatorname{Pr}(I_{t-1}=o, I_{t}=p \mid  \boldsymbol{\psi}_{t}) \boldsymbol{\theta}_{t \mid t}^{(o,p)}}{ \operatorname{Pr}(I_{t}=p \mid  \boldsymbol{\psi}_{t}) } 
        $,         \smallbreak
        \noindent
        $
            \boldsymbol{P}_{t \mid t}^{(p)} 
            = \frac{\sum_{o=0}^{1} \operatorname{Pr}(I_{t-1}=o, I_{t}=p \mid  \boldsymbol{\psi}_{t}) \left\{ \boldsymbol{P}_{t \mid t}^{(o,p)} + \left( \boldsymbol{\theta}_{t \mid t}^{(o,p)}-\boldsymbol{\theta}_{t \mid t}^{(p)} \right) \left( \boldsymbol{\theta}_{t \mid t}^{(o,p)}-\boldsymbol{\theta}_{t \mid t}^{(p)} \right)^T \right\}    }{ \operatorname{Pr}(I_{t}=p \mid  \boldsymbol{\psi}_{t}) }
        $.

        \item Collapse again to get marginal distribution \\
        $\boldsymbol{\theta}_{t \mid t} = \sum_{p=0}^{1} \operatorname{Pr}(I_{t}=p \mid \boldsymbol{\psi}_{t}) \boldsymbol{\theta}_{t \mid t}^{(p)}$,
        \\
        $\boldsymbol{P}_{t \mid t} = 
        \sum_{p=0}^{1} \operatorname{Pr}(I_{t}=p \mid \boldsymbol{\psi}_{t}) \left\{ \boldsymbol{P}_{t \mid t}^{(p)} + \left( \boldsymbol{\theta}_{t \mid t}^{(p)} - \boldsymbol{\theta}_{t \mid t}  \right) \left( \boldsymbol{\theta}_{t \mid t}^{(p)} - \boldsymbol{\theta}_{t \mid t}  \right)^T \right\}$.
    \end{enumerate}
\end{theorem}

There are two ways for appropriate initialization. We can either assume the initial states follow a multivariate Gaussian distribution with fixed but unknown mean and variance-covariance structure \citep{kim_dynamic_1994}, or assume it has a diffuse prior \citep{guo_structural_2000}. In addition, we can assume the initial probability $\operatorname{Pr}(I_{0}=o \mid \boldsymbol{\psi}_{0})$ with $o \in \{0,1\}$ is either a given value or a parameter to be estimated.
For missing values, there are two cases: (1) the entire observation vector is missing, or (2) part of the observation vector is missing. 
The MKF can deal with both cases using the details provided in Appendix\ref{de_mkf}.

The denominator in the second equation of step \ref{like} in Theorem \ref{MSSFS_MKF}, $p(\boldsymbol{y}_t | \boldsymbol{\psi}_{t-1})$, provides the likelihood of $\boldsymbol{y}_t$ given previous observations. This is used to compute the maximum likelihood estimate (MLE) based on equation \eqref{MSSFS_likelihood_approx}. Since observations from different subjects are independent, we can perform MKF for each subject and add their log-likelihood for the joint log-likelihood. Regularized estimates may be computed with a penalty to the likelihood. For example, we compute ridge estimates for some parameters in the real data analysis in Section \ref{MSSFS_RDA} since  some variables are highly correlated.

\subsection{Multiprocess Fixed Interval Smoothing (MFIS)} \label{MSSFS_estimation_MFIS}
While MKF aims primarily at parameter estimation and real-time detection, the smoothing algorithm complements filtering by providing a posterior estimate of the latent process. MKF operates in a forward manner, updating parameters and detecting changes as new data points are observed. In contrast, MFIS operates backward, starting from the last observed time point $n$ and refining estimates of the latent variables by incorporating information from the entire observation period. The details are presented in Theorem \ref{MSSFS_MFIS}.

\begin{theorem} \label{MSSFS_MFIS}
    Starting with $\boldsymbol{\theta}_{n \mid n}$, $\boldsymbol{P}_{n \mid n}$, and $\operatorname{Pr}(I_{n} \mid  \boldsymbol{\psi}_{n})$ from the last iteration of MKF, the MFIS with $I_t \in \{0,1\}$ consists of the following steps.  \\
     For $t=n-1,\ldots,1 $, DO:
    \begin{enumerate}
        \item Calculate the conditional mean and variance  \label{MFIS_1}
        \begin{flalign*}
        \boldsymbol{\theta}_{t+1 \mid t} &= \mathbb{E} \left( \boldsymbol{\theta}_{t+1} \mid \boldsymbol{\psi}_{t} \right) =\sum_{p=0}^{1} \sum_{q=0}^{1} \boldsymbol{\theta}_{t+1 \mid t}^{(p,q)} \operatorname{Pr}(I_t = p, I_{t+1} = q \mid \boldsymbol{\psi}_{t}), && \\
        \boldsymbol{\Sigma}_{t , t+1} &= 
        \operatorname{Cov} (\boldsymbol{\theta}_{t},\boldsymbol{\theta}_{t+1} \mid \boldsymbol{\psi}_{t}) && \\
        &= \sum_{p=0}^{1} \sum_{q=0}^{1} \operatorname{Pr}(I_t = p, I_{t+1} = q \mid \boldsymbol{\psi}_{t}) \left[ \boldsymbol{\theta}_{t|t}^{(p)} \boldsymbol{\gamma}_{q,t+1}^T + \boldsymbol{\theta}_{t|t}^{(p)} {\boldsymbol{\theta}_{t|t}^{(p)}}^T \mathbf{G}_{q,t+1}^T + \boldsymbol{P}_{t|t}^{(p)}  \mathbf{G}_{q,t+1}^T  \right] && \\
        &- \boldsymbol{\theta}_{t \mid t} \boldsymbol{\theta}_{t+1 \mid t}^T, && \\
        \boldsymbol{P}_{t+1 \mid t} 
        &= 
        \operatorname{Cov} (\boldsymbol{\theta}_{t+1} \mid \boldsymbol{\psi}_{t}) \\
        &=  
        \sum_{p=0}^{1} \sum_{q=0}^{1} \operatorname{Pr}(I_t = p, I_{t+1} = q \mid \boldsymbol{\psi}_{t}) \left[  \mathbf{G}_{q,t+1} \boldsymbol{P}_{t|t}^{(p)} \mathbf{G}_{q,t+1}^T + \mathbf{W}_{q,t+1} \right] \\
        &+
        \sum_{p=0}^{1} \sum_{q=0}^{1} \operatorname{Pr}(I_t = p, I_{t+1} = q \mid \boldsymbol{\psi}_{t}) \left[ \boldsymbol{\gamma}_{q,t+1} \boldsymbol{\gamma}_{q,t+1}^T + \boldsymbol{\gamma}_{q,t+1} {\boldsymbol{\theta}_{t|t}^{(p)}}^T \mathbf{G}_{q,t+1}^T \right.\\
        & \left. {}+ \mathbf{G}_{q,t+1} \boldsymbol{\theta}_{t|t}^{(p)} \boldsymbol{\gamma}_{q,t+1}^T + \mathbf{G}_{q,t+1} \boldsymbol{\theta}_{t|t}^{(p)} {\boldsymbol{\theta}_{t|t}^{(p)}}^T \mathbf{G}_{q,t+1}^T \right] \\
        &-
        \left(\sum_{p=0}^{1} \sum_{q=0}^{1} \operatorname{Pr}(I_t = p, I_{t+1} = q \mid \boldsymbol{\psi}_{t}) \left[ \boldsymbol{\gamma}_{q,t+1} + \mathbf{G}_{q,t+1} \boldsymbol{\theta}_{t|t}^{(p)} \right] \right) \\
        & \left( \sum_{p=0}^{1} \sum_{q=0}^{1} \operatorname{Pr}(I_t = p, I_{t+1} = q \mid \boldsymbol{\psi}_{t}) \left[ \boldsymbol{\gamma}_{q,t+1} + \mathbf{G}_{q,t+1} \boldsymbol{\theta}_{t|t}^{(p)} \right]\right)^T. &&
        \end{flalign*}

        \item Calculate conditional distributions \\
        $ \mathbb{E} \left(\boldsymbol{\theta}_t \mid \boldsymbol{\theta}_{t+1}, \psi_t \right)
        =
        \boldsymbol{\theta}_{t|t} + \boldsymbol{\Sigma}_{t,t+1} \boldsymbol{P}_{t+1|t}^{-1} (\boldsymbol{\theta}_{t+1}-\boldsymbol{\theta}_{t+1|t})
        $,\\   
        $\operatorname{Cov} \left(\boldsymbol{\theta}_t \mid \boldsymbol{\theta}_{t+1}, \psi_t \right)
        =
        \boldsymbol{P}_{t|t} - \boldsymbol{\Sigma}_{t,t+1} \boldsymbol{P}_{t+1|t}^{-1} \boldsymbol{\Sigma}_{t,t+1}^T
        $.

        \item Calculate marginal distributions \\
        $    \boldsymbol{\theta}_{t \mid T} =
            \boldsymbol{\theta}_{t|t} + \boldsymbol{\Sigma}_{t,t+1} \boldsymbol{P}_{t+1|t}^{-1} ( \boldsymbol{\theta}_{t+1|T}  -\boldsymbol{\theta}_{t+1|t})
        $,\\
        $    \boldsymbol{P}_{t \mid T} 
            = 
            \boldsymbol{P}_{t|t} - \boldsymbol{\Sigma}_{t,t+1} \boldsymbol{P}_{t+1|t}^{-1} \boldsymbol{\Sigma}_{t,t+1}^T + 
            \boldsymbol{\Sigma}_{t,t+1} \boldsymbol{P}_{t+1|t}^{-1} \boldsymbol{P}_{t+1|T}
            \boldsymbol{P}_{t+1|t}^{-1} \boldsymbol{\Sigma}_{t,t+1}^T
        $.

        \item Calculate the probability of interest \label{MFIS_4}\\
        $    \operatorname{Pr}(I_{t}=p, I_{t+1}=q \mid  \boldsymbol{\psi}_{T}) = \frac{\operatorname{Pr}(I_{t+1}=q \mid  \boldsymbol{\psi}_{T}) \operatorname{Pr}(I_{t}=p \mid   \boldsymbol{\psi}_{t}) \operatorname{Pr}(I_{t+1}=q \mid I_{t}=p,   \boldsymbol{\psi}_{t})}{\sum_{p=0}^{1} \operatorname{Pr}(I_{t}=p, I_{t+1}=q \mid \boldsymbol{\psi}_{t})}
        $, \\
        $
        \operatorname{Pr}(I_{t}=p \mid  \boldsymbol{\psi}_{T}) = \sum_{q=0}^{1} \operatorname{Pr}(I_{t}=p,I_{t+1}=q \mid \boldsymbol{\psi}_{T})
        $.
    \end{enumerate}
\end{theorem}

The proof of the theorem is provided in Appendix \ref{de_mfis}.

\section{Modeling Temperature Changes in COVID-19-Infected Hemodialysis Patients} \label{MSSFS_RDA}
The proposed MSSFS enables both real-time change detection using MKF and the investigation of feedback systems using MFIS. This section utilizes the proposed methods to examine temperature changes in hemodialysis (HD) patients infected with COVID-19. 

We consider a dataset of 43 HD patients who received in-center treatment at Fresenius Medical Care and tested positive for COVID-19 via reverse transcription polymerase chain reaction (RT-PCR) between January 1, 2020, and August 31, 2021. We align temperature measurements from different patients to the RT-PCR test date, designated as day $0$, and include measurements taken 50 days before and 50 days after this date. 
We refer to these observations as the positive arm. For the negative arm, for each patient, we randomly select a 101-day observation window before January 1, 2020. The average number of temperature observations per patient per arm is $41$. We include two covariates for the switching probabilities: gender and age. Gender is encoded as $1$ for males and $0$ for females, with males representing $60.5$\% of all patients. The average age of patients is $62.55$ years, with a standard deviation of $10.75$ years. We standardized the age covariate for each patient in each arm to have a mean of $0$ and a standard deviation of $1$.

The system equations characterize how the underlying change in temperature ($\theta_{i,t}$) evolves under different statuses. 
We will consider system equations motivated by the thermoregulation models \citep{boulant_neuronal_2006}. Based on existing literature, a synaptic network of hypothalamic neurons is responsible for a set-point thermoregulation model where four types of neurons, warm-sensitive, temperature-insensitive, heat loss effector neuron, heat production effector neuron, regulate the body temperature change from one level to another. In our real data analysis, we focus on COVID-19-related fever. We assume each patient has two possible statuses at time $t$: $I_{t}=0$ (no fever) and $I_{t}=1$ (fever). The temperature is assumed to be at reference level $k$ (ref$_k$) when $I_{t}=k$ for $k=0,1$. The change in body temperature at time $t$ is proportional to the difference between the previous body temperature and the reference level, with a rate and random disturbance of $r_k$ and $w_{t,k}$, respectively, when $I_{t}=k$ for $k=0,1$. The system equations can be represented as follows:
\begin{align}
    &\text{System Equations:} &\theta_{t} &= \left(\theta_{t-1}-\text{ref}_0\right) r_0 + \theta_{t-1} + w_{t,0} \quad \text{if } I_{t}=0,  \nonumber \\
    &  &\theta_{t} &= \left(\theta_{t-1}-\text{ref}_1\right) r_1 + \theta_{t-1} + w_{t,1} \quad \text{if }I_{t}=1, \nonumber
\end{align}
where $\theta_{t}$ represent the body temperature at time $t$, and $r_0$ and $r_1$ are the change rates. $r_0$ and $r_1$ take values in $[-1,0)$. A value close to zero means the current temperature does not change much and remains at the previous value; a value close to $-1$ indicates the temperature changes quickly to the corresponding level. A value in between indicates the temperature gradually changes from the previous value to the corresponding reference level.
 

Since normal body temperature varies by individual and we focus on temperature changes rather than absolute body temperature, we consider the deviation from the no-fever status as the response. Specifically, we define the change in temperature as observed body temperature minus temperature when there is no fever, $\text{ref}_0$. We estimate $\text{ref}_0$ for each patient as their average body temperature based on measurements taken from days $-100$ to $-51$. Consequently, we consider the following MSSFS:

\begin{flalign}
&\text{Observation Equation:} &y_{i,t} &= \theta_{i,t} + v_t,  \label{TOE_2} && \\
&\text{System Equation:} &\theta_{i,t} &= G_{0} \theta_{i,t-1} + w_{t,0} \quad \text{if } I(t)=0,   \label{TSE1_2}  &&  \\
&  &\theta_{i,t} &=  \gamma_{1} + G_{1} \theta_{i,t-1} + w_{t,1} \quad \text{if } I(t)=1, \label{TSE2_2} && \\
&\text{Switch Equation:} & \pi_{i,t,01} &= \frac{\exp \{\alpha_0 + \boldsymbol{x}_i^T \boldsymbol{\beta}_0\}}{1+\exp\{\alpha_0 + \boldsymbol{x}_i^T \boldsymbol{\beta}_0 \}}, \label{TSWE1} \\
&                        & \pi_{i,t,11} &= \frac{\exp \{\alpha_1 + \boldsymbol{x}_i^T \boldsymbol{\beta}_1 +  \zeta_{1} \boldsymbol{w}^T \boldsymbol{\theta}_{i,(t-L):(t-1)}  \}}{1+\exp \{\alpha_1 + \boldsymbol{x}_i^T \boldsymbol{\beta}_1 +  \zeta_{1} \boldsymbol{w}^T \boldsymbol{\theta}_{i,(t-L):(t-1)}  \}}, \label{TSWE2}
\end{flalign}
where $y_{i,t}$ is the change in temperature at time $t$ from patient $i$, $\theta_{i,t}$ is the underlying change in temperature, $G_0 = (1+r_0) \in [0,1)$ and $G_1 = (1+r_1) \in [0,1)$, $\delta = \text{ref}_1 - \text{ref}_0 = \text{ref}_1 > 0$ is the average temperature rise due to fever, and $\gamma_1 = -\delta r_1$.
We assume that $v_t \stackrel{iid}{\sim} N(0,\sigma_{v}^2)$ for all $t$, $w_0 \stackrel{iid}{\sim} N(0,\sigma_0^2)$, and $w_1 \stackrel{iid}{\sim} N(0,\sigma_1^2)$.



For the switch equations, we assume that the transition probability from no fever to fever ($\pi_{i,t,01}$) does not depend on the history of the temperature change since fever is usually caused by external factors rather than the temperature change itself. On the other hand, we assume that the transition probability $\pi_{i,t,11}$ depends on the previous temperature changes since a higher temperature (fever) in the past could affect how the body temperature status changes in the future. 
We assume $\pi_{i,t,11}$ depends on an exponentially weighted average of the previous $L$ states $\boldsymbol{w}^T\boldsymbol{\theta}_{i,(t-L):(t-1)}$, where $\boldsymbol{w}=(w_1,\cdots,w_L)^T$, $w_l = e^{\rho (L-l+1)}$ for $l=1,\cdots,L$, and $\rho<0$ is a decay rate.
Considering the small sample size in this analysis, instead of estimating $\rho$ as a parameter, we set $\rho=0.5$. We tested other choices of $\rho$, and the results remain similar. Using notations in the switching equations in the general MSSFS model \eqref{SE3} and \eqref{SE4},
we set $z_{i,t,0}(\boldsymbol{\theta}_{i,1:(t-1)})=0$ and $z_{i,t,1}(\boldsymbol{\theta}_{i,1:(t-1)})=\zeta_{1} \boldsymbol{w}^T\boldsymbol{\theta}_{i,(t-L):(t-1)}$. 
These choices lead to better and more interpretable results than other functions of $z_{i,t,k}$ we have considered. This formulation also has a similar interpretation as the short-term cumulative
effects used in the distributed lag model \citep{welty_bayesian_2009}. 

Denote $\boldsymbol{\Theta} = (\sigma_v^2,\sigma_0^2,\sigma_1^2,\delta, G_0, G_1,\alpha_0,\boldsymbol{\beta}_0^T,\alpha_1,\boldsymbol{\beta}_1^T,\zeta_1)^T$ as the collection of all parameters in the above model, where
$\boldsymbol{\beta}_k = (\beta_{male,k},\beta_{age,k})^T$ for $k=0,1$ are the coefficients of covariates in the switch equations. All variances and $\delta$ are estimated using natural log transformation, and $G_0$ and $G_1$ are estimated using a logit transformation to enforce constraints. From the preliminary study, we set $L=3$ in the switch equation. Other values lead to similar conclusions. 
To start the MKF, we need to initialize the conditional mean ($\boldsymbol{\theta}_{0 \mid 0}^{(o)}$), conditional variance ($\boldsymbol{P}_{0 \mid 0}^{(o)}$), and the initial probability ($\operatorname{Pr}(I_{0}=o \mid \boldsymbol{\psi}_{0})$) for $o \in \{0,1\}$. In this study, we assume all patients start with a status of no fever, i.e., $I_0=0$. Thus, $\boldsymbol{\theta}_{0 \mid 0}^{(0)} = 0$, $\boldsymbol{\theta}_{0 \mid 0}^{(1)} = \delta$, $\boldsymbol{P}_{0 \mid 0}^{(0)} = \boldsymbol{P}_{0 \mid 0}^{(1)} = 0$, and $\operatorname{Pr}(I_{0}=0 \mid \boldsymbol{\psi}_{0}) = 1-\operatorname{Pr}(I_{0}=1 \mid \boldsymbol{\psi}_{0}) = 1$.

We observed that the estimates of $\alpha_1$ and $\zeta_1$ are highly correlated since the estimated change in temperature among close time points are highly correlated. To obtain more stable estimates of parameters in the switch probabilities, we consider penalized likelihood with an $L_2$ penalty to $\alpha_1$, $\boldsymbol{\beta}_1$, and $\zeta_1$. We set the tuning parameter as  $0.01$, and results with other choices of the tuning parameter are similar.
The confidence intervals are constructed using bootstrap with 300 repetitions. We employ the bias-corrected and accelerated (BCa) method to create bootstrap confidence intervals, addressing potential bias in the $L_2$ penalized estimates \citep{efron_introduction_1993}.
We use the limited-memory Broyden–Fletcher–Goldfarb–Shanno (L-BFGS-B) algorithm \citep{byrd_limited_1995,zhu_algorithm_1997} to compute the update of parameters. We set the parameter in the EM algorithm as follows: $\mathcal{N}_{max}=30$, $D_{EM} = 0.001$ and $\kappa = 10^{-6}$. 

\begin{figure}[H]
    \centering
    \includegraphics[width=\textwidth]{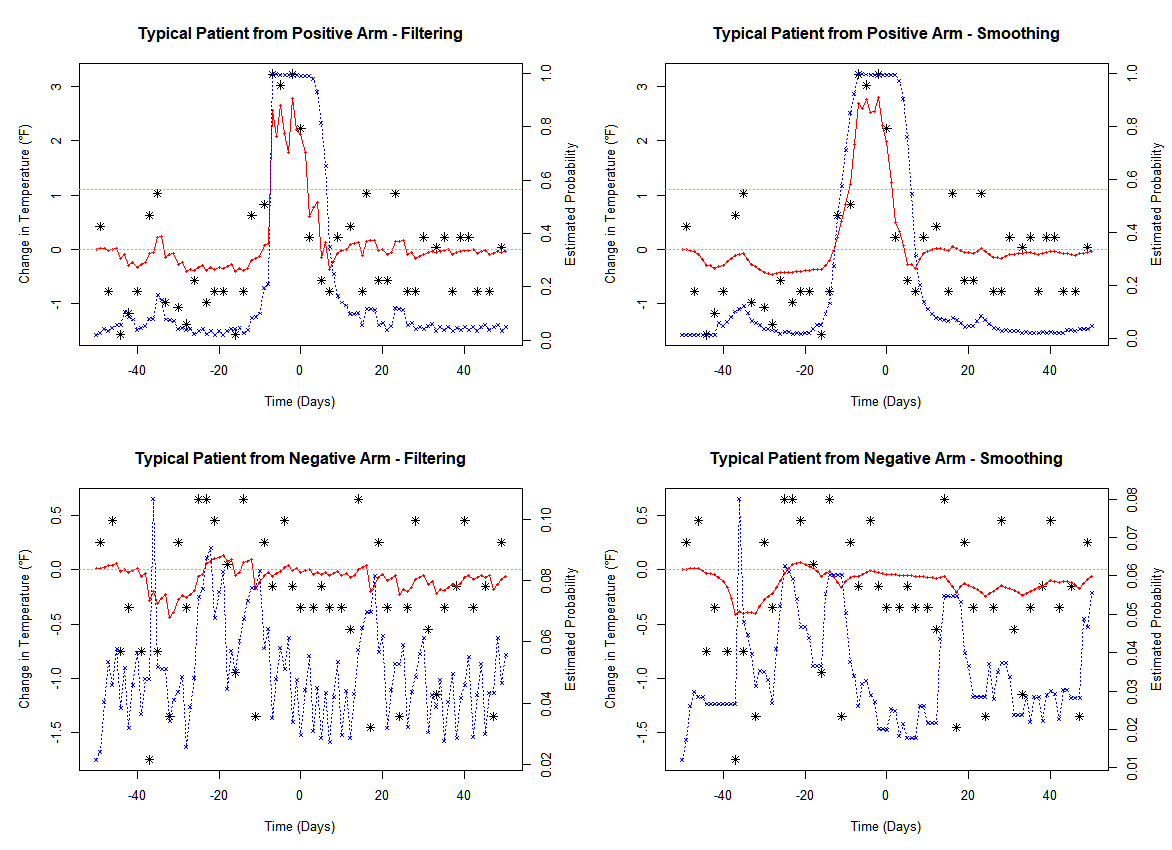}
    \caption{Estimate of states and probabilities of having a fever for one typical patient in both arms. The upper and lower panels correspond to positive and negative arms, while the left and right panels present filtering and smoothing estimates. The black star points are the observed change in temperatures. The red solid line with dots is the filtering (left panels) and smoothing (right panels) estimate of the underlying change in temperature, calculated as $\mathbb{E}(\boldsymbol{\theta}_{t} \mid \boldsymbol{\psi}_t)$ (left panels) and $\mathbb{E}(\boldsymbol{\theta}_{t} \mid \boldsymbol{\psi}_T)$ (right panels) for $t = 1, \ldots, n$. The blue dashed line with crosses is the filtering (left panels) and smoothing (right panels) estimate of probabilities for having a fever, calculated as $\operatorname{Pr}(I_{t} = 1 \mid \boldsymbol{\psi}_{t})$ (left panels) and $\operatorname{Pr}(I_{t} = 1 \mid \boldsymbol{\psi}_{T})$ (right panels) for $t = 1, \ldots, n$. The two orange horizontal lines indicate $0$ and $\hat{\delta}$, respectively. Only one orange line appears in each of the bottom plots, as the line corresponding to $\hat{\delta}$ falls outside the displayed range.}
    \label{MSSFS_figRD1}
\end{figure}

Table \ref{MSSFS_Tab:RD} presents the estimates of parameters and their 95\% bootstrap confidence intervals. The estimated average temperature change is $1.1$ degrees Fahrenheit with a 95\% confidence interval $(0.95,1.35)$. The deacreasing rate $r_0 = G_0 - 1$ is small while increasing rate $r_1 = G_1 - 1$ is relatively larger. This can be seen from Figure \ref{MSSFS_figRD1} where the temperature increases faster than it decreases.

\begin{table}
\caption{Estimates of parameters and 95\% confidence intervals.}
\centering
\begin{tabular}{ V{3} c|c|c V{3} } 
\specialrule{.2em}{.1em}{.1em}
Parameters & Estimate & 95\% CI  \\
\hline 
$\sigma_v^2$&  0.3839  &(0.3684, 0.4440) \\
\hline
$\sigma_0^2$&  0.0000  &(0.0000, 0.3327)\\
\hline
$\sigma_1^2$&  0.6218  &(0.3699, 0.9128) \\
\hline
$\delta$&  1.0979  &(0.9618, 1.3491) \\
\hline
$G_0$&  0.9392  &(0.9371, 0.9583)  \\
\hline
$G_1$& 0.6651   &(0.4416, 0.7488)  \\
\hline
$\alpha_0$&  -3.7057   &(-4.7902, -3.0090)  \\
\hline
$\beta_{male,0}$&  0.1616    &(-0.5471, 7.1694)  \\
\hline
$\beta_{age,0}$&   -0.1896  &(-1.0063, -0.0205)  \\
\hline
$\alpha_1$&  1.0606  &(-1.8209, 3.6355) \\
\hline
$\beta_{male,1}$&   -0.8571   &(-3.2938, -0.2788)  \\
\hline
$\beta_{age,1}$&  0.8377   &(0.6461, 1.3878)  \\
\hline
$\zeta_1$&   2.5922  &(0.7363, 5.9954) \\
\specialrule{.2em}{.1em}{.1em}
\end{tabular}
\label{MSSFS_Tab:RD}
\end{table}

For the parameters in the switching probabilities, a positive estimate of $\beta_{male,0}$, though not statistically significant, suggests that male patients are more likely to develop a fever after a COVID-19 infection, aligning with previous findings \citep{patel_gender_2023}. A negative estimate of $\beta_{age,0}$ indicates that older patients are less likely to develop a fever after infection, which is also consistent with prior research \citep{ma_nonparametric_2024}.
A negative estimate of $\beta_{male,1}$ suggests that male patients are less likely to recover from a fever after a COVID-19 infection, while a positive estimate of $\beta_{age,1}$ indicates that older patients are more likely to recover from a fever. Similar findings for patients over $40$ have been reported in the literature \citep{halimatuzzahro_effect_2023, voinsky_effects_2020}. We note that all patients in our study are older than $40$.

Figure \ref{MSSFS_figRD1} illustrates the estimates of underlying states and the probabilities of having a fever for one patient in both the positive and negative arms. Plots for other patients show similar patterns. The proposed methods yield effective filtering and smoothing estimates of the underlying states and transition probabilities, which can be utilized for online detection of COVID-19 infection.


\begin{figure}[H]
    \centering
    \includegraphics[width=\textwidth]{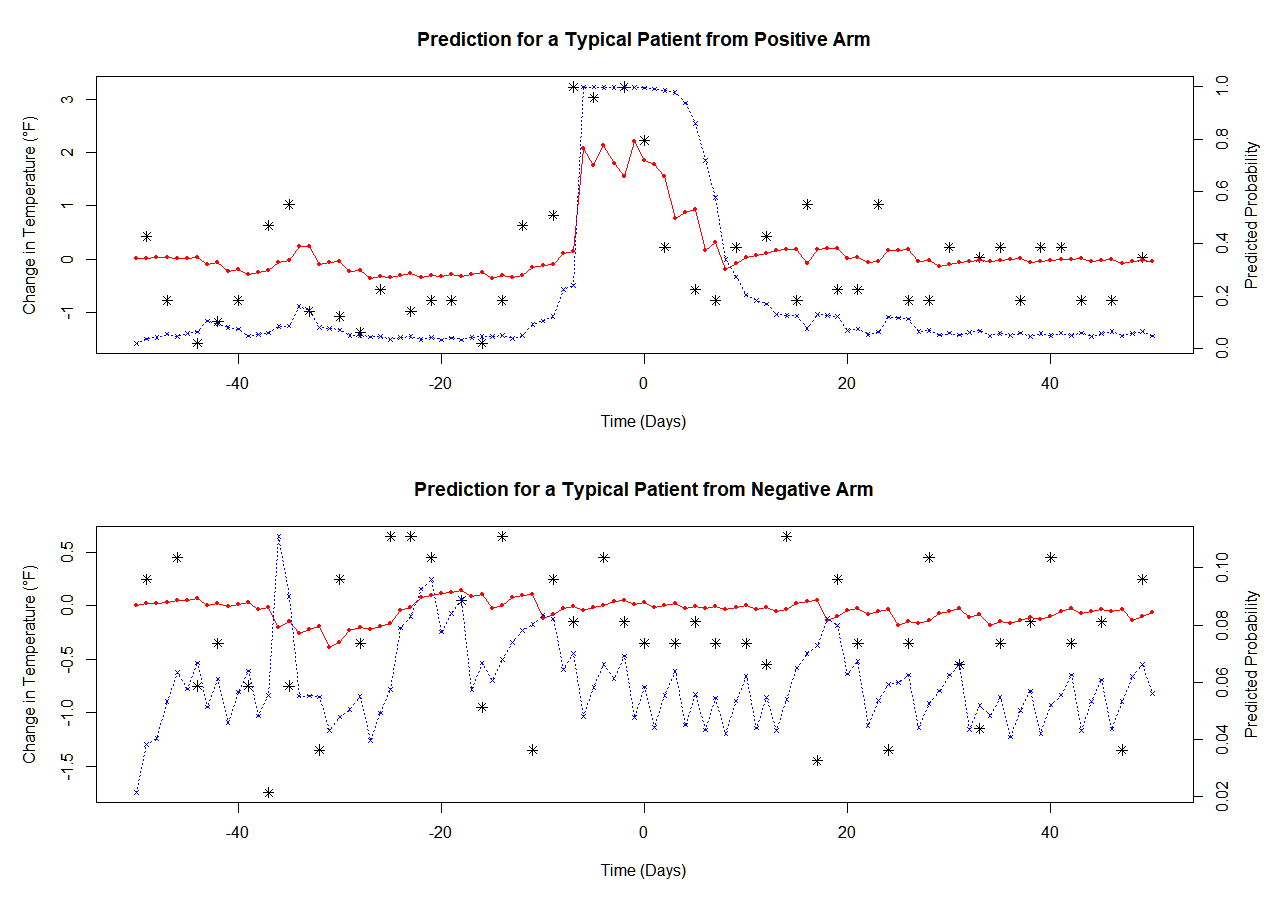}
    \caption{The one-step forward prediction of the change in underlying temperature and the probability of having a fever for the same typical patient in the main text. The upper panel shows the prediction of the patient where time 0 corresponds to the positive PCR test date. The lower panel shows the prediction of the same patient in the negative arm, whose observations were randomly selected during the COVID-free period. The black star points are the observed change in temperatures from the baseline. The red solid line with dots is the prediction for the change in underlying body temperature from the baseline, which is calculated using $\mathbb{E}(\boldsymbol{\theta}_{t+1} \mid \boldsymbol{\psi}_t)$ for $t = 0, \ldots, n-1$, which formula can be found in MFIS step (1). The blue dashed line with crosses is the predicted probability of having a fever, which is calculated using $\operatorname{Pr}(I_{t+1} = 1 \mid \boldsymbol{\psi}_{t})$ for $t = 0, \ldots, n-1$, which formula can be found in the denominator of MFIS step (4).}
    \label{MSSFS_figRD2}
\end{figure}

The proposed method also provides predictions of the latent states. As an illustration, Figure \ref{MSSFS_figRD2} shows the one-step forward prediction of the change in underlying temperature and the probability of having a fever for the same typical patient in the main text. 
The plots demonstrate that the proposed model provides reasonable good predictions and is highly responsive to temperature changes, with the probability increasing as the temperature rises.

\section{Simulation Study for MSSFS} \label{MSSFS_sim}

We conduct simulations to evaluate the accuracy of parameter estimates and the time complexity across various scenarios. 
We consider the same model equations (\ref{TOE_2}) to (\ref{TSE2_2}). 
To imitate the real data analysis, we consider two covariates for the transition probabilities, $x_{i1}$ and $x_{i2}$, with parameters $\boldsymbol{\beta}_k=(\beta_{k1},\beta_{k2})^T$ for $k=1,2$. 
We generate the covariates $x_{i1}$ and $x_{i2}$ similar to the real data: $x_{i1}$ is a binary variable, taking the value 1 with a probability of 60.5\%, while $x_{i2}$ is a continuous variable generated from a standard Gaussian distribution. The lag in the switch equation is set to $L=3$, as in the real data analysis. For other parameters, we set $\sigma_v^2 = 0.1$, $\sigma_0^2 = 0.03$, $\sigma_1^2 = 0.3$, $r_0 = -0.5$, $r_1 = -0.5$, $\alpha_0 = -3$, $\beta_{01}=0.15$, $\beta_{02} = -0.2$, $\beta_{11} = -0.8$, and $\beta_{12} = 0.5$. We consider two choices of $\delta$, $\delta=5$ and $\delta=10$, for two different signal-to-noise levels. To evaluate the effect of sample size on the estimates, we consider three sample sizes for the number of subjects:  $m=100$, $300$, and $500$. For different feedback mechanisms, we consider two scenarios: one for positive feedback with $\alpha_1 = 0.2$ and $\zeta_1 = 0.3$, and another for negative feedback with $\alpha_1 = 4$ and $\zeta_1 = -0.3$. This results in a total of $12$ simulation settings. 

To check whether the algorithm is robust to the initial value of parameters, we set the initial values of parameters arbitrarily as follows: all the variances are set to 1, and all parameters in the switch equations are set to 0. We do not apply the $L_2$ penalty for the simulation studies, and we set the maximum number of iterations and convergence criteria for the EM algorithm to be the same as in real data analysis.

\begin{figure}[H]
    \centering
    \includegraphics[width=\textwidth]{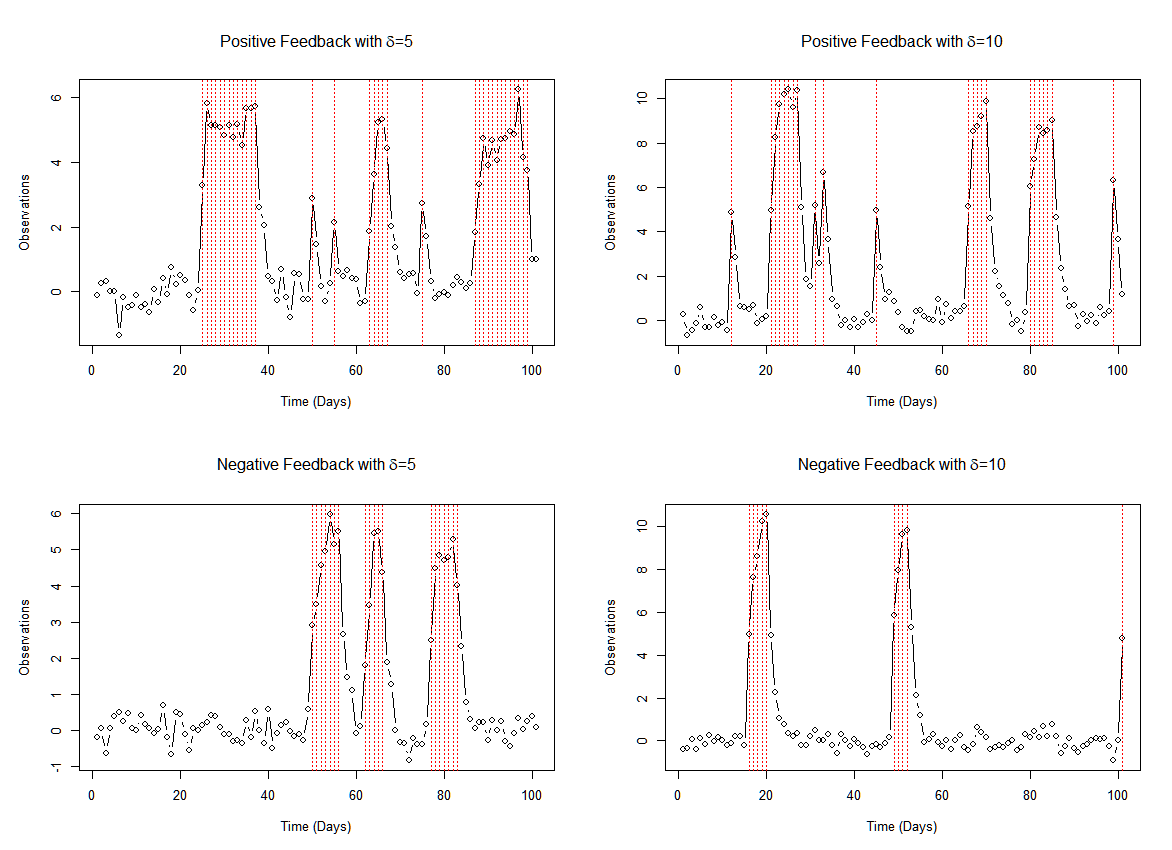}
    \caption{Plots of observations under different signal-to-noise levels and feedback mechanisms. The circles are the responses, and the verticle red dashed lines indicate times at status $I(t)=1$.}
    \label{MSSFS_illustrate}
\end{figure}

Figure \ref{MSSFS_illustrate} demonstrates how different signal-to-noise ratios and feedback patterns influence the observations of an individual. Observations are cleaner under higher signal-to-noise ratios and exhibit more regular change patterns in the negative feedback setting. The estimated positive feedback in the real data suggests that fever patterns vary across individuals. In our simulations, we consider both positive and negative feedback since both may arise in real-world applications.

Tables \ref{MSSFS_Tab:sim_p_1} to \ref{MSSFS_Tab:sim_n_2} present mean squared error (MSE), squared bias (Bias$^2$), and variance (Var) of all 13 parameters under different simulation settings.  
We separate the parameters into two groups: parameters in the observation and state equations and parameters in the switch equations. Tables \ref{MSSFS_Tab:sim_p_1}, \ref{MSSFS_Tab:sim_p_2} present the results for the positive feedback, and Tables \ref{MSSFS_Tab:sim_n_1}, \ref{MSSFS_Tab:sim_n_2} present the result for the negative feedback. Overall, despite involving several approximations, the proposed estimation method produces good estimates for all parameters. The accuracy improves with higher signal-to-noise ratios and/or larger sample sizes. 

\begin{table}[H]
\caption{MSEs, biases, and variances, all multiplied by 100, for parameters in the observation and state equations with positive feedback.}
\centering
\begin{tabular}{ V{3} c|c|c|c|c|c|c|c V{3} } 
\specialrule{.2em}{.1em}{.1em}
\multicolumn{2}{V{3} c|}{} &  \multicolumn{3}{c|}{$\delta$=5} & \multicolumn{3}{c V{3}}{$\delta$=10}\\ 
\hline 
Parameters & Metrics & m=100 & m=300 & m=500 & m=100 & m=300 & m=500\\ 
\hline 
\multirow{3}{*}{$\sigma_{v}^2$}  & MSE & 0.0011 & 0.0003 & 0.0002 & 0.0012 & 0.0004 & 0.0002 \\ 
& Bias$^2$ & 0.0000 & 0.0000 & 0.0000 & 0.0000 & 0.0000 & 0.0000 \\ 
& Variance & 0.0011 & 0.0003 & 0.0002 & 0.0012 & 0.0004 & 0.0002 \\ 
\hline 
\multirow{3}{*}{$\sigma_{0}^2$}  & MSE & 0.0006 & 0.0002 & 0.0001 & 0.0008 & 0.0002 & 0.0001 \\ 
& Bias$^2$ & 0.0000 & 0.0000 & 0.0000 & 0.0000 & 0.0000 & 0.0000 \\ 
& Variance & 0.0006 & 0.0002 & 0.0001 & 0.0008 & 0.0002 & 0.0001 \\ 
\hline 
\multirow{3}{*}{$\sigma_{1}^2$}  & MSE & 0.0290 & 0.0090 & 0.0046 & 0.0103 & 0.0035 & 0.0022 \\ 
& Bias$^2$ & 0.0007 & 0.0004 & 0.0001 & 0.0001 & 0.0001 & 0.0000 \\ 
& Variance & 0.0283 & 0.0086 & 0.0045 & 0.0102 & 0.0035 & 0.0022 \\ 
\hline 
\multirow{3}{*}{$\delta$}  & MSE & 0.2170 & 0.0671 & 0.0382 & 0.0445 & 0.0189 & 0.0126 \\ 
& Bias$^2$ & 0.0046 & 0.0013 & 0.0007 & 0.0001 & 0.0001 & 0.0001 \\ 
& Variance & 0.2124 & 0.0658 & 0.0375 & 0.0444 & 0.0188 & 0.0125 \\ 
\hline 
\multirow{3}{*}{$G_{0}$}  & MSE & 0.0013 & 0.0005 & 0.0003 & 0.0004 & 0.0001 & 0.0001 \\ 
& Bias$^2$ & 0.0001 & 0.0001 & 0.0001 & 0.0001 & 0.0000 & 0.0000 \\ 
& Variance & 0.0012 & 0.0004 & 0.0002 & 0.0003 & 0.0001 & 0.0001 \\ 
\hline 
\multirow{3}{*}{$G_{1}$}  & MSE & 0.0058 & 0.0019 & 0.0012 & 0.0011 & 0.0004 & 0.0003 \\ 
& Bias$^2$ & 0.0001 & 0.0001 & 0.0000 & 0.0000 & 0.0000 & 0.0000 \\ 
& Variance & 0.0057 & 0.0018 & 0.0012 & 0.0011 & 0.0004 & 0.0003 \\ 
\specialrule{.2em}{.1em}{.1em}
\end{tabular}
\label{MSSFS_Tab:sim_p_1}
\end{table}

\begin{table}[H]
\caption{MSEs, biases, and variances, all multiplied by 100, for parameters in the switch equations with positive feedback.}
\centering
\begin{tabular}{ V{3} c|c|c|c|c|c|c|c V{3} } 
\specialrule{.2em}{.1em}{.1em}
\multicolumn{2}{V{3} c|}{} &  \multicolumn{3}{c|}{$\delta$=5} & \multicolumn{3}{c V{3}}{$\delta$=10}\\ 
\hline 
Parameters & Metrics & m=100 & m=300 & m=500 & m=100 & m=300 & m=500\\ 
\hline 
\multirow{3}{*}{$\alpha_{0}$} & MSE & 0.7733 & 0.2547 & 0.1608 & 1.0284 & 0.3365 & 0.2051 \\ 
& Bias$^2$ & 0.0127 & 0.0002 & 0.0000 & 0.0079 & 0.0001 & 0.0000 \\ 
& Variance & 0.7606 & 0.2545 & 0.1608 & 1.0205 & 0.3364 & 0.2051 \\ 
\hline 
\multirow{3}{*}{$\alpha_{1}$} & MSE & 3.3626 & 1.8206 & 1.1802 & 3.6297 & 1.1275 & 0.7657 \\ 
& Bias$^2$ & 0.3266 & 0.5932 & 0.4778 & 0.0063 & 0.0222 & 0.0449 \\ 
& Variance & 3.0360 & 1.2274 & 0.7024 & 3.6234 & 1.1053 & 0.7208 \\ 
\hline 
\multirow{3}{*}{$\beta_{01}$} & MSE & 1.0608 & 0.4007 & 0.2216 & 1.2845 & 0.4868 & 0.2837 \\ 
& Bias$^2$ & 0.0087 & 0.0001 & 0.0001 & 0.0182 & 0.0000 & 0.0000 \\ 
& Variance & 1.0521 & 0.4006 & 0.2215 & 1.2663 & 0.4868 & 0.2837 \\ 
\hline 
\multirow{3}{*}{$\beta_{02}$} & MSE & 0.2414 & 0.0840 & 0.0471 & 0.2844 & 0.0997 & 0.0497 \\ 
& Bias$^2$ & 0.0000 & 0.0002 & 0.0009 & 0.0002 & 0.0005 & 0.0016 \\ 
& Variance & 0.2414 & 0.0838 & 0.0462 & 0.2842 & 0.0992 & 0.0481 \\ 
\hline 
\multirow{3}{*}{$\beta_{11}$} & MSE & 1.4096 & 0.5566 & 0.3168 & 1.8098 & 0.5893 & 0.2713 \\ 
& Bias$^2$ & 0.0005 & 0.0222 & 0.0039 & 0.0823 & 0.0300 & 0.0112 \\ 
& Variance & 1.4091 & 0.5344 & 0.3129 & 1.7275 & 0.5593 & 0.2601 \\ 
\hline 
\multirow{3}{*}{$\beta_{12}$} & MSE & 0.5922 & 0.1803 & 0.1191 & 0.5146 & 0.1775 & 0.0993 \\ 
& Bias$^2$ & 0.0065 & 0.0146 & 0.0060 & 0.0042 & 0.0019 & 0.0001 \\ 
& Variance & 0.5857 & 0.1657 & 0.1131 & 0.5104 & 0.1756 & 0.0992 \\ 
\hline 
\multirow{3}{*}{$\zeta_{1}$} & MSE & 0.2187 & 0.1207 & 0.0865 & 0.0395 & 0.0144 & 0.0105 \\ 
& Bias$^2$ & 0.0233 & 0.0430 & 0.0401 & 0.0002 & 0.0012 & 0.0012 \\ 
& Variance & 0.1954 & 0.0777 & 0.0464 & 0.0393 & 0.0132 & 0.0093 \\
\specialrule{.2em}{.1em}{.1em}
\end{tabular}
\label{MSSFS_Tab:sim_p_2}
\end{table}

\begin{table}[H]
\caption{MSEs, biases, and variances, all multiplied by 100, for parameters in the observation and state equations with negative feedback.}
\centering
\begin{tabular}{ V{3} c|c|c|c|c|c|c|c V{3} } 
\specialrule{.2em}{.1em}{.1em}
\multicolumn{2}{V{3} c|}{} &  \multicolumn{3}{c|}{$\delta$=5} & \multicolumn{3}{c V{3}}{$\delta$=10}\\
\hline 
Parameters & Metrics & m=100 & m=300 & m=500 & m=100 & m=300 & m=500\\  
\hline 
\multirow{3}{*}{$\sigma_{v}^2$} & MSE & 0.0014 & 0.0004 & 0.0002 & 0.0011 & 0.0003 & 0.0002 \\ 
& Bias$^2$ & 0.0000 & 0.0000 & 0.0000 & 0.0000 & 0.0000 & 0.0000 \\ 
& Var & 0.0014 & 0.0004 & 0.0002 & 0.0011 & 0.0003 & 0.0002 \\ 
\hline 
\multirow{3}{*}{$\sigma_{0}^2$} & MSE & 0.0008 & 0.0003 & 0.0001 & 0.0006 & 0.0002 & 0.0001 \\ 
& Bias$^2$ & 0.0000 & 0.0000 & 0.0000 & 0.0000 & 0.0000 & 0.0000 \\ 
& Var & 0.0008 & 0.0003 & 0.0001 & 0.0006 & 0.0002 & 0.0001 \\ 
\hline 
\multirow{3}{*}{$\sigma_{1}^2$} & MSE & 0.0117 & 0.0035 & 0.0019 & 0.0184 & 0.0053 & 0.0029 \\ 
& Bias$^2$ & 0.0003 & 0.0001 & 0.0000 & 0.0002 & 0.0001 & 0.0000 \\ 
& Var & 0.0114 & 0.0034 & 0.0019 & 0.0182 & 0.0052 & 0.0029 \\ 
\hline 
\multirow{3}{*}{$\delta$} & MSE & 0.0463 & 0.0142 & 0.0076 & 0.1826 & 0.0559 & 0.0306 \\ 
& Bias$^2$ & 0.0004 & 0.0002 & 0.0001 & 0.0002 & 0.0010 & 0.0010 \\ 
& Var & 0.0459 & 0.0140 & 0.0075 & 0.1824 & 0.0549 & 0.0296 \\ 
\hline 
\multirow{3}{*}{$G_{0}$} & MSE & 0.0011 & 0.0004 & 0.0003 & 0.0003 & 0.0001 & 0.0000 \\ 
& Bias$^2$ & 0.0001 & 0.0000 & 0.0001 & 0.0000 & 0.0000 & 0.0000 \\ 
& Var & 0.0010 & 0.0004 & 0.0002 & 0.0003 & 0.0001 & 0.0000 \\ 
\hline 
\multirow{3}{*}{$G_{1}$} & MSE & 0.0039 & 0.0011 & 0.0009 & 0.0013 & 0.0005 & 0.0003 \\ 
& Bias$^2$ & 0.0000 & 0.0000 & 0.0000 & 0.0000 & 0.0000 & 0.0000 \\ 
& Var & 0.0039 & 0.0011 & 0.0009 & 0.0013 & 0.0005 & 0.0003 \\
\specialrule{.2em}{.1em}{.1em}
\end{tabular}
\label{MSSFS_Tab:sim_n_1}
\end{table}

\begin{table}[H]
\caption{MSEs, biases, and variances, all multiplied by 100, for parameters in the switch equations with negative feedback.}
\centering
\begin{tabular}{ V{3} c|c|c|c|c|c|c|c V{3} } 
\specialrule{.2em}{.1em}{.1em}
\multicolumn{2}{V{3} c|}{} &  \multicolumn{3}{c|}{$\delta$=5} & \multicolumn{3}{c V{3}}{$\delta$=10}\\
\hline 
Parameters & Metrics & m=100 & m=300 & m=500 & m=100 & m=300 & m=500\\ 
\hline 
\multirow{3}{*}{$\alpha_{0}$} & MSE & 0.8686 & 0.3297 & 0.1918 & 0.7488 & 0.2662 & 0.1562 \\ 
& Bias$^2$ & 0.0109 & 0.0013 & 0.0000 & 0.0124 & 0.0021 & 0.0005 \\ 
& Variance & 0.8577 & 0.3284 & 0.1918 & 0.7364 & 0.2641 & 0.1557 \\ 
\hline 
\multirow{3}{*}{$\alpha_{1}$} & MSE & 11.6267 & 6.9949 & 6.7991 & 5.8496 & 2.3164 & 1.4914 \\ 
& Bias$^2$ & 4.4079 & 4.8575 & 5.4261 & 0.0385 & 0.1818 & 0.1770 \\ 
& Variance & 7.2188 & 2.1374 & 1.3730 & 5.8111 & 2.1346 & 1.3144 \\ 
\hline 
\multirow{3}{*}{$\beta_{01}$} & MSE & 1.0905 & 0.5107 & 0.2726 & 0.9317 & 0.3813 & 0.2038 \\ 
& Bias$^2$ & 0.0043 & 0.0012 & 0.0001 & 0.0066 & 0.0013 & 0.0003 \\ 
& Variance & 1.0862 & 0.5095 & 0.2725 & 0.9251 & 0.3800 & 0.2035 \\ 
\hline 
\multirow{3}{*}{$\beta_{02}$} & MSE & 0.2760 & 0.1015 & 0.0502 & 0.2464 & 0.0836 & 0.0484 \\ 
& Bias$^2$ & 0.0000 & 0.0001 & 0.0013 & 0.0003 & 0.0004 & 0.0011 \\ 
& Variance & 0.2760 & 0.1014 & 0.0489 & 0.2461 & 0.0832 & 0.0473 \\ 
\hline 
\multirow{3}{*}{$\beta_{11}$} & MSE & 1.5780 & 0.5101 & 0.3086 & 1.6530 & 0.5702 & 0.3164 \\ 
& Bias$^2$ & 0.0073 & 0.0032 & 0.0085 & 0.0067 & 0.0037 & 0.0062 \\ 
& Variance & 1.5707 & 0.5069 & 0.3001 & 1.6463 & 0.5665 & 0.3102 \\ 
\hline 
\multirow{3}{*}{$\beta_{12}$} & MSE & 0.3628 & 0.1361 & 0.0737 & 0.3635 & 0.0994 & 0.0668 \\ 
& Bias$^2$ & 0.0470 & 0.0194 & 0.0148 & 0.0005 & 0.0001 & 0.0010 \\ 
& Variance & 0.3158 & 0.1167 & 0.0589 & 0.3630 & 0.0993 & 0.0658 \\ 
\hline 
\multirow{3}{*}{$\zeta_{1}$} & MSE & 0.5059 & 0.3383 & 0.3173 & 0.0723 & 0.0326 & 0.0203 \\ 
& Bias$^2$ & 0.2515 & 0.2492 & 0.2627 & 0.0006 & 0.0024 & 0.0020 \\ 
& Variance & 0.2544 & 0.0891 & 0.0546 & 0.0717 & 0.0302 & 0.0183 \\ 
\specialrule{.2em}{.1em}{.1em}
\end{tabular}
\label{MSSFS_Tab:sim_n_2}
\end{table}

\begin{figure}[H]
    \centering    \includegraphics[width=0.7\textwidth]{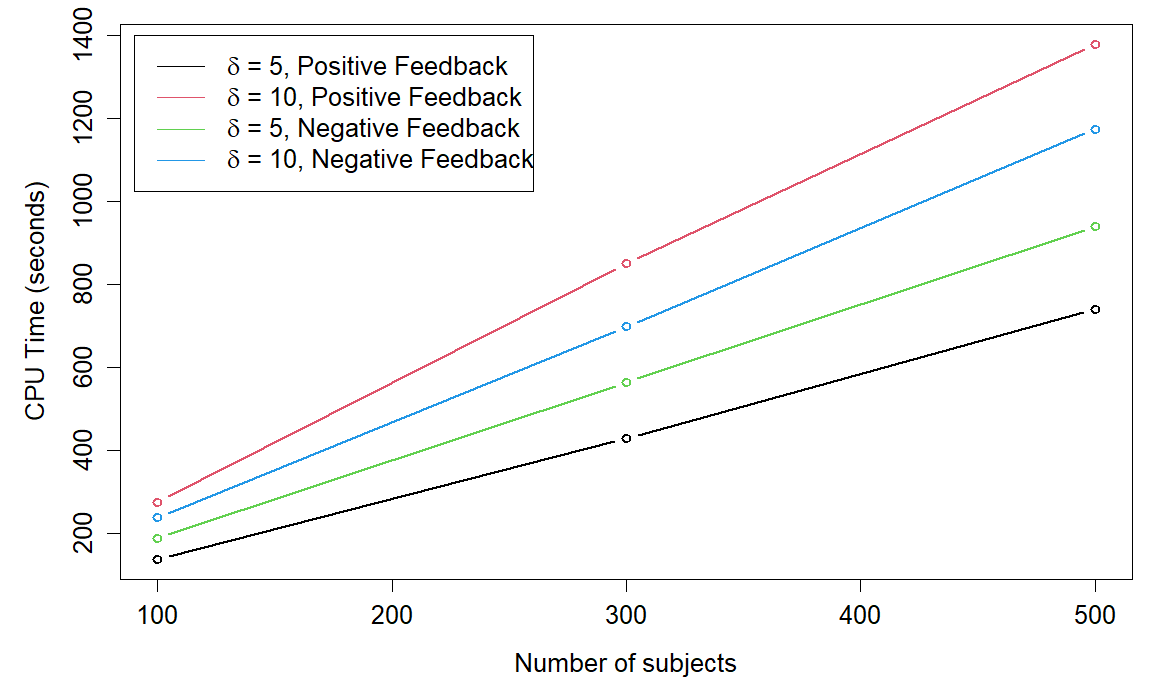}
    \caption{Plot of average CPU time per simulation versus the number of subjects under different settings.}
    \label{MSSFS_fig_time}
\end{figure}

Figure \ref{MSSFS_fig_time} shows that CPU time increases linearly with the number of subjects. This indicates that the proposed estimation procedure retains a key advantage of Kalman filtering with linear time complexity. Such linear scalability makes our model especially well-suited for handling large datasets.

\section{Conclusions} \label{conclusion}

We propose a multiprocess state space model to study dynamics across different health conditions, introducing a flexible feedback mechanism in the transition probabilities between these conditions. An EM algorithm is developed, incorporating extended MKF and MFIS to estimate latent states and parameters. Efficient approximations are derived to achieve linear time complexity. Simulation studies demonstrate that the proposed estimation method performs well. We apply the model to analyze temperature dynamics in COVID-19-infected hemodialysis patients, with several findings aligning with existing literature.

The proposed model and methods can be adapted to other models and applications. For instance, some authors have studied mixtures of AR models \citep{wong_mixture_2000, boshnakov_analytic_2009}, and incorporating these into a multiprocess state space model could be a promising direction. Additionally, modern research explores combining state space models with neural networks \citep{zamarreno_state_1998, van_lint_accurate_2005, liu_predicting_2006}. Further research is needed to investigate the potential of integrating a multiprocess state space model with neural networks to capture different underlying processes.

\section*{Acknowledgements}

This research is partially supported by NIH grants R01-DK130067.

\appendix 
\section{Derivation of MKF} \label{de_mkf}
In this section, we provide the derivation of the Multiprocess Kalman Filter (MKF) in Theorem 1. The italicized text outlines approaches for handling missing values in two scenarios: (a) when all elements of $\boldsymbol{y}_t$ are missing, and (b) when only some elements of $\boldsymbol{y}_t$ are missing.

\begin{proof}
\text{ }
    \begin{enumerate}
        \item Initialize 
        $\boldsymbol{\theta}_{t-1 \mid t-1}^{(o)} = \mathbb{E}\left[\boldsymbol{\theta}_{t-1} \mid I_{t-1}=o, \boldsymbol{\psi}_{t-1}\right]$ and
        $\boldsymbol{P}_{t-1 \mid t-1}^{(o)} = \operatorname{Cov}\left[\boldsymbol{\theta}_{t-1} \mid I_{t-1}=o, \boldsymbol{\psi}_{t-1}\right]$ when $t=1$. 
        There are two common approaches in literature: either regard $\boldsymbol{\theta}_{0 \mid 0}^{(o)}$ and $\boldsymbol{P}_{0 \mid 0}^{(o)}$ for $o \in \{0,1\}$ as fixed but unknown parameters to be estimated \citep{kim_dynamic_1994} or use a diffuse distribution for the initial state vector \citep{guo_structural_2000}. In this paper, we regard them as fixed but unknown parameters. 

        \item \label{MKF2_proof}
        Calculate one step forward prediction as follows:
        \begin{flalign*}
            \boldsymbol{\theta}_{t \mid t-1}^{(o,p)} 
            &= \mathbb{E}\left[ \boldsymbol{\theta}_{t} \mid \boldsymbol{\psi}_{t-1} ,I_{t-1}=o, I_{t}=p \right] \\
            &= \mathbb{E}\left[ \boldsymbol{\gamma}_{I_t,t} + \mathbf{G}_{I_t,t} \boldsymbol{\theta}_{t-1} + \boldsymbol{w}_{I_t,t} \mid \boldsymbol{\psi}_{t-1} ,I_{t-1}=o, I_{t}=p \right] \\
            &= \boldsymbol{\gamma}_{p,t} + \mathbf{G}_{p,t} \boldsymbol{\theta}_{t-1 \mid t-1}^{(o)}, && \\
            \boldsymbol{P}_{t \mid t-1}^{(o,p)} 
            &= \operatorname{Cov}\left( \boldsymbol{\gamma}_{I_t,t} + \mathbf{G}_{I_t,t} \boldsymbol{\theta}_{t-1} + \boldsymbol{w}_{I_t,t} \mid \boldsymbol{\psi}_{t-1} ,I_{t-1}=o, I_{t}=p \right) \\
            &= \mathbf{G}_{p,t}  \boldsymbol{P}_{t-1 \mid t-1}^{(o)}  \mathbf{G}_{p,t}^T + \mathbf{W}_{pt}, && \\
            \boldsymbol{y}_{t \mid t-1}^{(o,p)} 
            &= \mathbf{F}_t \boldsymbol{\theta}_{t \mid t-1}^{(o,p)}, &&
        \end{flalign*}

    \bigbreak
    \noindent
    where the last equations in the conditional mean and covariance use the fact that $\boldsymbol{\theta}_{t-1}$ and $I_t$ are independent given $I_{t-1}$, $\boldsymbol{x}_i$ and $\hat{\boldsymbol{z}}_{1:n}^{\tau-1}$ at the previous EM iteration.
    
 {\it 
    To deal with missing data in $\boldsymbol{y}_t$,  
    \begin{itemize}
        \item If all elements of $\boldsymbol{y}_t$ are missing, 
         skip the last equation above since it will not be used in the next step. 
        \item If some elements of $\boldsymbol{y}_t$ are missing, 
        we define a subsetting matrix $\mathbf{S}_t$ by removing the rows corresponding to the indices of missing elements of an identity matrix. Multiplying the subsetting matrix on both sides of the observation equation, we have
        \begin{align*}
            \boldsymbol{y}^{*}_t 
            =
            \mathbf{F}^{*}_t \boldsymbol{\theta}_t+\boldsymbol{v}^{*}_t,
        \end{align*}
        where $\boldsymbol{y}^{*}_t = \mathbf{S}_t \boldsymbol{y}_t$ contains the observed elements, $\mathbf{F}^{*}_t = \mathbf{S}_t \mathbf{F}_t$, and 
        $\boldsymbol{v}^{*}_t = \mathbf{S}_t \boldsymbol{v}_t \sim
        N\left(\boldsymbol{0},  \mathbf{V}^{*}_t  \right)
        =
        N\left(\boldsymbol{0}, \mathbf{S}_t \mathbf{V}_t \mathbf{S}_t^T \right)$. 
        
        With this new equation, the last equation in step \ref{MKF2_proof} becomes
        $
        \boldsymbol{y}_{t \mid t-1}^{(o,p)*} 
        = \mathbf{F}_t^{*} \boldsymbol{\theta}_{t \mid t-1}^{(o,p)}.
        $
    \end{itemize}
}
        
        \bigbreak
        \item \label{MKF3_proof} 
        Calculate one-step observation prediction error and prediction error variance:
        \begin{flalign*}
        \boldsymbol{\eta}_{t \mid t-1}^{(o,p)} &= \boldsymbol{y}_t - \boldsymbol{y}_{t \mid t-1}^{(o,p)}, && \\
        \mathbf{H}_{t \mid t-1}^{(o,p)} &= \operatorname{Cov}(\boldsymbol{\eta}_{t \mid t-1}^{(o,p)} \mid \boldsymbol{\psi}_{t-1} ,I_{t-1}=o, I_{t}=p) &&\\
        &= \operatorname{Cov}(\mathbf{F}_{t} \boldsymbol{\theta}_t + \boldsymbol{v}_t - \mathbf{F}_{t} \boldsymbol{\theta}_{t \mid t-1}^{(o,p)}  \mid \boldsymbol{\psi}_{t-1} ,I_{t-1}=o, I_{t}=p) &&\\
        &= \operatorname{Cov}(\mathbf{F}_{t} \boldsymbol{\theta}_t + \boldsymbol{v}_t \mid \boldsymbol{\psi}_{t-1} ,I_{t-1}=o, I_{t}=p) &&\\
        &= \mathbf{F}_{t} \boldsymbol{P}_{t \mid t-1}^{(o,p)} \mathbf{F}_{t}^T + \mathbf{V}_t. &&
        \end{flalign*}

    {\it
    To deal with missing data in $\boldsymbol{y}_t$,  
    \begin{itemize}
        \item If all elements of $\boldsymbol{y}_t$ are missing, skip this step since this step will not contribute to the log-likelihood.
        \item If some elements of $\boldsymbol{y}_t$ is missing, use the following formula instead: \\
            $
            \boldsymbol{\eta}_{t \mid t-1}^{(o,p)*} = \boldsymbol{y}_t^{*} - \boldsymbol{y}_{t \mid t-1}^{(o,p)*} 
            $ ,\\
            $\mathbf{H}_{t \mid t-1}^{(o,p)*}
            =
            \mathbf{S}_{t} \mathbf{F}_{t} \boldsymbol{P}_{t \mid t-1}^{(o,p)} \mathbf{F}_{t}^T \mathbf{S}_{t}^T+ \mathbf{S}_{t} \mathbf{V}_t \mathbf{S}_{t}^T
            $.
    \end{itemize}
    }

        \bigbreak
        \item \label{MKF4_proof}
        Calculate the posterior mean of states and their variances $\boldsymbol{\theta}_{t \mid t}^{(o,p)}$ and $\boldsymbol{P}_{t \mid t}^{(o,p)}$ using the following approximate joint distribution to a mixture of Gaussian distributions \citep{kim_dynamic_1994}:
        \bigbreak
        $
        \left(\begin{array}{l}
        \boldsymbol{\theta}_t \\
        \boldsymbol{y}_t
        \end{array} \bigg\rvert \boldsymbol{\psi}_{t-1}, I_{t-1}=o, I_t = p\right)
        \stackrel{\text{approx}}{\sim}        \operatorname{MN}\left(
        \begin{bmatrix}
        \boldsymbol{\theta}_{t \mid t-1}^{(o,p)} \\
        \boldsymbol{y}_{t \mid t-1}^{(o,p)} 
        \end{bmatrix},
        \begin{bmatrix}
        \boldsymbol{P}_{t \mid t-1}^{(o,p)}  & \boldsymbol{P}_{t \mid t-1}^{(o,p)} \mathbf{F}_{t}^T\\
        \mathbf{F}_{t} \boldsymbol{P}_{t \mid t-1}^{(o,p)}  & \mathbf{F}_{t} \boldsymbol{P}_{t \mid t-1}^{(o,p)} \mathbf{F}_{t}^T + \mathbf{V}_t
        \end{bmatrix}
        \right).
    $
    \bigbreak
    Using the conditional distribution formulae for multivariate Gaussian distributions, we have

\begin{flalign*}
    \boldsymbol{\theta}_{t \mid t}^{(o,p)} 
    &= \mathbb{E}[\boldsymbol{\theta}_{t}  \mid  \boldsymbol{\psi}_{t}, I_{t-1}=o, I_{t}=p ] & &\\
    &= \boldsymbol{\theta}_{t \mid t-1}^{(o,p)} + \boldsymbol{P}_{t \mid t-1}^{(o,p)} \mathbf{F}_{t}^T \left( \mathbf{F}_{t} \boldsymbol{P}_{t \mid t-1}^{(o,p)} \mathbf{F}_{t}^T + \mathbf{V}_t \right)^{-1} \boldsymbol{\eta}_{t \mid t-1}^{(o,p)}, && \\
    \boldsymbol{P}_{t \mid t}^{(o,p)} 
    &= \operatorname{Cov}[\boldsymbol{\theta}_{t}  \mid  \boldsymbol{\psi}_{t}, I_{t-1}=o, I_{t}=p ] && \\
    &= \boldsymbol{P}_{t \mid t-1}^{(o,p)} - \boldsymbol{P}_{t \mid t-1}^{(o,p)} \mathbf{F}_{t}^T \left( \mathbf{F}_{t} \boldsymbol{P}_{t \mid t-1}^{(o,p)} \mathbf{F}_{t}^T + \mathbf{V}_t \right)^{-1} \mathbf{F}_{t} \boldsymbol{P}_{t \mid t-1}^{(o,p)}. &&
\end{flalign*}

    {\it
    To deal with missing data in $\boldsymbol{y}_t$,  
    \begin{itemize}
        \item If all elements of $\boldsymbol{y}_t$ are missing, replace the above formula with the following:\\
             $
            \boldsymbol{\theta}_{t \mid t}^{(o,p)} 
            = \mathbb{E}[\boldsymbol{\theta}_{t}  \mid  \boldsymbol{\psi}_{t}, I_{t-1}=o, I_{t}=p ] 
            = 
            \mathbb{E}[\boldsymbol{\theta}_{t}  \mid  \boldsymbol{\psi}_{t-1}, I_{t-1}=o, I_{t}=p ]
            =
            \boldsymbol{\theta}_{t \mid t-1}^{(o,p)} ,$
            \bigbreak
            $
                \boldsymbol{P}_{t \mid t}^{(o,p)} 
                = \operatorname{Cov}[\boldsymbol{\theta}_{t}  \mid  \boldsymbol{\psi}_{t}, I_{t-1}=o, I_{t}=p ]
                = \operatorname{Cov}[\boldsymbol{\theta}_{t}  \mid  \boldsymbol{\psi}_{t-1}, I_{t-1}=o, I_{t}=p ]
                = \boldsymbol{P}_{t \mid t-1}^{(o,p)}.
            $
        \item If some elements of $\boldsymbol{y}_t$ is missing, replace $\boldsymbol{y}_t$,
            $\boldsymbol{y}_{t \mid t-1}^{(o,p)}$,
            $\mathbf{F}_{t}$, and
            $\mathbf{V}_t$ with 
            $\boldsymbol{y}^{*}_t$,
            $\boldsymbol{y}_{t \mid t-1}^{(o,p)*}$,
            $\mathbf{F}^{*}_t$, and
            $\mathbf{V}^{*}_t$.
    \end{itemize}
    }


        \bigbreak
        \item \label{MKFprob_proof} To finish MKF and then compute the MLE, we need to update probabilities $\operatorname{Pr}(I_{t-1}=o, I_{t}=p \mid \boldsymbol{\psi}_t)$ and $\operatorname{Pr}(I_{t}=p \mid \boldsymbol{\psi}_t)$.
        
        \begin{enumerate}
            \item Start with $\operatorname{Pr}(I_{t-1}=o \mid \boldsymbol{\psi}_{t-1})$. 
            
            At $t=1$, we could either initialize this probability as a given value or set it as an unknown parameter. In this paper, we assume all patients start without a fever and have $\operatorname{Pr}(I_{0}=0 \mid \boldsymbol{\psi}_{0}) = 1$.
            
            \bigbreak
            \item \label{MKFprob1_proof}
        $
            \operatorname{Pr}(I_{t-1}=o, I_{t}=p \mid \boldsymbol{\psi}_{t-1}) = \operatorname{Pr}(I_{t}=p \mid I_{t-1}=o,  \boldsymbol{\psi}_{t-1}) \operatorname{Pr}(I_{t-1}=o \mid \boldsymbol{\psi}_{t-1})
        $.
        
            $\operatorname{Pr}(I_{t}=p \mid I_{t-1}=o,  \boldsymbol{\psi}_{t-1})$ is known since it is the transition probability which depends on the covariates and feedback $\boldsymbol{z}_{1:(t-1)}$. Here we use the estimate $\hat{\boldsymbol{z}}_{1:(t-1)}$ from the previous step. When $t<L+1$, the transition probability $\operatorname{Pr}(I_{t}=p \mid I_{t-1}=o,  \boldsymbol{\psi}_{t-1})$ depends on the first a few lags only. For example, if $L=3$ and $t=2$, only the system's history at $t=1$ is used. If a weighted average is used, then the weights are reweighted according to their original values.

        \bigbreak
        
            \bigbreak
            \item \label{MKFprob2_proof}
        $
            p(\boldsymbol{y}_t, I_{t-1}=o, I_{t}=p \mid \boldsymbol{\psi}_{t-1}) = p(\boldsymbol{y}_t \mid I_{t-1}=o, I_{t}=p,  \boldsymbol{\psi}_{t-1})  \operatorname{Pr}(I_{t-1}=o, I_{t}=p \mid \boldsymbol{\psi}_{t-1})
        $.
        
            The first term on the right equals 
            $(2 \pi)^{-n/2} |\mathbf{H}_{t \mid t-1}^{(o,p)}|^{-1/2} \exp(-\frac{1}{2} {\boldsymbol{\eta}_{t|t-1}^{(o,p)}}^T (\mathbf{H}_{t \mid t-1}^{(o,p)})^{-1} \boldsymbol{\eta}_{t|t-1}^{(o,p)})$, and the second term on the right comes from step \ref{MKFprob1_proof}.
            
    {\it
    To deal with missing data in $\boldsymbol{y}_t$,  
    \begin{itemize}
        \item If all elements of $\boldsymbol{y}_t$ are missing, skip this step.
        \item If some elements of $\boldsymbol{y}_t$ is missing, the first term on the right is replaced by \\
            $
                (2 \pi)^{-n/2} |\mathbf{H}_{t \mid t-1}^{(o,p)*}|^{-1/2} \exp(-\frac{1}{2} {\boldsymbol{\eta}_{t|t-1}^{(o,p)*}}^T (\mathbf{H}_{t \mid t-1}^{(o,p)*})^{-1} \boldsymbol{\eta}_{t|t-1}^{(o,p)*}).
            $
    \end{itemize}
    }


            \item \label{MKFprob3_proof}
            \begin{flalign*} 
                \operatorname{Pr}(I_{t-1}=o, I_{t}=p \mid \boldsymbol{\psi}_{t}) 
                &= \frac{p(\boldsymbol{y}_t, I_{t-1}=o, I_{t}=p \mid \boldsymbol{\psi}_{t-1})}{p(\boldsymbol{y}_t \mid \boldsymbol{\psi}_{t-1})} && \\
                &= \frac{p(\boldsymbol{y}_t, I_{t-1}=o, I_{t}=p \mid \boldsymbol{\psi}_{t-1})}{\sum_{p=0}^{1} \sum_{o=0}^{1} p(\boldsymbol{y}_t, I_{t-1}=o, I_{t}=p \mid \boldsymbol{\psi}_{t-1})}.  &&
            \end{flalign*}
            \bigbreak
            The numerator and denominator are from step (\ref{MKFprob2_proof}).

{\it
    To deal with missing data in $\boldsymbol{y}_t$,  
    \begin{itemize}
        \item If all elements of $\boldsymbol{y}_t$ are missing, follow the same step as in step (\ref{MKFprob1_proof}) since we do not have observation at time $t$, \\
            $
             \operatorname{Pr}(I_{t-1}=o, I_{t}=p \mid \boldsymbol{\psi}_{t}) 
             =
             \operatorname{Pr}(I_{t-1}=o, I_{t}=p \mid \boldsymbol{\psi}_{t-1})
            $.
        \item If some elements of $\boldsymbol{y}_t$ is missing, use the alternative formula from step (\ref{MKFprob2_proof}).
    \end{itemize}
    }

            \item  \label{MKFprob4_proof}
        $
            \operatorname{Pr}(I_{t}=p \mid \boldsymbol{\psi}_{t}) = \sum_{o=0}^{1} \operatorname{Pr}(I_{t-1}=o, I_{t}=p \mid \boldsymbol{\psi}_{t}) 
        $.
        \bigbreak
        The result of step (\ref{MKFprob4_proof}) is the starting point for the next iteration of step \ref{MKFprob_proof}. The denominator of step (\ref{MKFprob3_proof}), $p(\boldsymbol{y}_t \mid \boldsymbol{\psi}_{t-1})$, is used for MLE. When we have multiple subjects, the MKF is run for each subject, and the log-likelihood is summed up for MLE.
        
{\it
    To deal with missing data in $\boldsymbol{y}_t$,  
    \begin{itemize}
        \item If all elements of $\boldsymbol{y}_t$ are missing, the same formula with replacements from the previous step. No need to update the log-likelihood.
        \item If some elements of $\boldsymbol{y}_t$ are missing, use the substituted formula from step (\ref{MKFprob3_proof}).
    \end{itemize}
}

        \end{enumerate}

        \item \label{MKFprob5_proof}
        Using the probabilities in step \ref{MKFprob1_proof} to \ref{MKFprob4_proof}, we can collapse the statuses by marginalizing out $I_{t-1}$,    
        \begin{align*}
        \boldsymbol{\theta}_{t \mid t}^{(p)} &=
        \mathbb{E}\left[\boldsymbol{\theta}_t \mid I_{t}=p, \boldsymbol{\psi}_t\right] \\
        &=
        \sum_{o=0}^{1} \operatorname{Pr}(I_{t-1}=o \mid I_{t}=p, \boldsymbol{\psi}_{t}) \boldsymbol{\theta}_{t \mid t}^{(o,p)}  \\
        &= \frac{\sum_{o=0}^{1} \operatorname{Pr}(I_{t-1}=o, I_{t}=p \mid  \boldsymbol{\psi}_{t}) \boldsymbol{\theta}_{t \mid t}^{(o,p)}}{ \operatorname{Pr}(I_{t}=p \mid  \boldsymbol{\psi}_{t}) }, \\
        \boldsymbol{P}_{t \mid t}^{(p)} &=\mathbb{E}\left[\left(\boldsymbol{\theta}_t-\mathbb{E}\left[\boldsymbol{\theta}_t \mid I_{t}=p, \boldsymbol{\psi}_t\right]\right)\left(\boldsymbol{\theta}_t-\mathbb{E}\left[\boldsymbol{\theta}_t \mid I_{t}=p, \boldsymbol{\psi}_t\right]\right)^T \mid I_{t}=p, \boldsymbol{\psi}_t\right] \\        &=\mathbb{E}\left[\left(\boldsymbol{\theta}_t-\boldsymbol{\theta}_{t \mid t}^{(p)}\right)\left(\boldsymbol{\theta}_t-\boldsymbol{\theta}_{t \mid t}^{(p)}\right)^T \mid I_{t}=p, \boldsymbol{\psi}_t\right] \\
        &=\sum_{o=0}^{1} \frac{\operatorname{Pr}(I_{t-1}=o, I_{t}=p \mid  \boldsymbol{\psi}_{t}) }{ \operatorname{Pr}(I_{t}=p \mid  \boldsymbol{\psi}_{t}) } \mathbb{E}\left[\left(\boldsymbol{\theta}_t-\boldsymbol{\theta}_{t \mid t}^{(p)}\right)\left(\boldsymbol{\theta}_t-\boldsymbol{\theta}_{t \mid t}^{(p)}\right)^T \mid I_{t-1}=o, I_{t}=p, \boldsymbol{\psi}_t\right] \\
        &=\sum_{o=0}^{1} \frac{\operatorname{Pr}(I_{t-1}=o, I_{t}=p \mid  \boldsymbol{\psi}_{t}) }{ \operatorname{Pr}(I_{t}=p \mid  \boldsymbol{\psi}_{t}) }, \\ 
        &{} \mathbb{E} \left[ \left(\boldsymbol{\theta}_t-\boldsymbol{\theta}_{t \mid t}^{(o,p)}+\boldsymbol{\theta}_{t \mid t}^{(o,p)}-\boldsymbol{\theta}_{t \mid t}^{(p)}\right)\left(\boldsymbol{\theta}_t-\boldsymbol{\theta}_{t \mid t}^{(o,p)}+\boldsymbol{\theta}_{t \mid t}^{(o,p)}-\boldsymbol{\theta}_{t \mid t}^{(p)}\right)^T \mid I_{t-1}=o, I_{t}=p, \boldsymbol{\psi}_t \right] \\
        &= \sum_{o=0}^{1} \frac{\operatorname{Pr}(I_{t-1}=o, I_{t}=p \mid  \boldsymbol{\psi}_{t}) }{ \operatorname{Pr}(I_{t}=p \mid  \boldsymbol{\psi}_{t}) }\left\{\mathbb{E}\left[\left(\boldsymbol{\theta}_t-\boldsymbol{\theta}_{t \mid t}^{(o,p)}\right)\left(\boldsymbol{\theta}_t-\boldsymbol{\theta}_{t \mid t}^{(o,p)}\right)^T \mid I_{t-1}=o, I_{t}=p, \boldsymbol{\psi}_t\right]\right.\\
        &\left.+\left(\boldsymbol{\theta}_{t \mid t}^{(p)}-\boldsymbol{\theta}_{t \mid t}^{(o,p)}\right)\left(\boldsymbol{\theta}_{t \mid t}^{(p)}-\boldsymbol{\theta}_{t \mid t}^{(o,p)}\right)^T\right\} \\
        &+\sum_{o=0}^{1} \frac{\operatorname{Pr}(I_{t-1}=o, I_{t}=p \mid  \boldsymbol{\psi}_{t}) }{ \operatorname{Pr}(I_{t}=p \mid  \boldsymbol{\psi}_{t}) }\left(\mathbb{E}\left[\boldsymbol{\theta}_t \mid I_{t-1}=o, I_{t}=p, \boldsymbol{\psi}_t\right]-\boldsymbol{\theta}_{t \mid t}^{(o,p)}\right)\left(\boldsymbol{\theta}_{t \mid t}^{(o,p)}-\boldsymbol{\theta}_{t \mid t}^{(p)}\right)^T \\
        &+\sum_{o=0}^{1} \frac{\operatorname{Pr}(I_{t-1}=o, I_{t}=p \mid  \boldsymbol{\psi}_{t}) }{ \operatorname{Pr}(I_{t}=p \mid  \boldsymbol{\psi}_{t}) }\left(\boldsymbol{\theta}_{t \mid t}^{(o,p)}-\boldsymbol{\theta}_{t \mid t}^{(p)}\right)\left(\mathbb{E}\left[\boldsymbol{\theta}_t \mid I_{t-1}=o, I_{t}=p, \boldsymbol{\psi}_t\right]-\boldsymbol{\theta}_{t \mid t}^{(o,p)}\right)^T\\
        &=\sum_{o=0}^{1} \frac{\operatorname{Pr}(I_{t-1}=o, I_{t}=p \mid  \boldsymbol{\psi}_{t}) }{ \operatorname{Pr}(I_{t}=p \mid  \boldsymbol{\psi}_{t}) }\left\{\mathbb{E}\left[\left(\boldsymbol{\theta}_t-\boldsymbol{\theta}_{t \mid t}^{(o,p)}\right)\left(\boldsymbol{\theta}_t-\boldsymbol{\theta}_{t \mid t}^{(o,p)}\right)^T \mid I_{t-1}=o, I_{t}=p, \boldsymbol{\psi}_t\right]\right. \\
        &\left.+\left(\boldsymbol{\theta}_{t \mid t}^{(p)}-\boldsymbol{\theta}_{t \mid t}^{(o,p)}\right)\left(\boldsymbol{\theta}_{t \mid t}^{(p)}-\boldsymbol{\theta}_{t \mid t}^{(o,p)}\right)^T\right\} \\
        &= \frac{\sum_{o=0}^{1} \operatorname{Pr}(I_{t-1}=o, I_{t}=p \mid  \boldsymbol{\psi}_{t}) \left\{ \boldsymbol{P}_{t \mid t}^{(o,p)} + \left( \boldsymbol{\theta}_{t \mid t}^{(o,p)}-\boldsymbol{\theta}_{t \mid t}^{(p)} \right) \left( \boldsymbol{\theta}_{t \mid t}^{(o,p)}-\boldsymbol{\theta}_{t \mid t}^{(p)} \right)^T \right\}    }{ \operatorname{Pr}(I_{t}=p \mid  \boldsymbol{\psi}_{t}) }.
        \end{align*}
        \bigbreak
        
        \item 
        Collapse again by marginalizing out $I_{t}$ to get marginal distribution
        \begin{flalign*}
        \boldsymbol{\theta}_{t \mid t} &=
        \mathbb{E}\left[\boldsymbol{\theta}_t \mid \boldsymbol{\psi}_t\right]
        =
        \sum_{p=0}^{1} \operatorname{Pr}(I_{t}=p \mid \boldsymbol{\psi}_{t}) \mathbb{E}\left[\boldsymbol{\theta}_t \mid I_{t}=p, \boldsymbol{\psi}_t\right] = 
        \sum_{p=0}^{1} \operatorname{Pr}(I_{t}=p \mid \boldsymbol{\psi}_{t}) \boldsymbol{\theta}_{t \mid t}^{(p)},  && \\
        \boldsymbol{P}_{t \mid t} &= 
        \sum_{p=0}^{1} \operatorname{Pr}(I_{t}=p \mid \boldsymbol{\psi}_{t}) \left\{ \boldsymbol{P}_{t \mid t}^{(p)} + \left( \boldsymbol{\theta}_{t \mid t}^{(p)} - \boldsymbol{\theta}_{t \mid t}  \right) \left( \boldsymbol{\theta}_{t \mid t}^{(p)} - \boldsymbol{\theta}_{t \mid t}  \right)^T \right\}.
        \end{flalign*}
        where the last equation follows the same step as in step (\ref{MKFprob5_proof}).

    \end{enumerate}

\end{proof}

\section{Derivation of MFIS} \label{de_mfis}
In this section, we provide the derivation of the Multiprocess Fixed Interval Smoothing (MFIS) in Theorem 2.
\begin{proof}
\text{ }
\begin{enumerate}
    \item Marginal distribution can be calculated by conditioning on $\boldsymbol{\theta}_{t+1}$:
    \begin{flalign*}
    \boldsymbol{\theta}_{t \mid T} &=
    \mathbb{E}\left[\boldsymbol{\theta}_t \mid \boldsymbol{\psi}_T\right]
    =
    \mathbb{E}\left[ \mathbb{E} \left(\boldsymbol{\theta}_t \mid \boldsymbol{\theta}_{t+1}, \psi_T \right) \mid \boldsymbol{\psi}_T\right] 
    =
    \mathbb{E}\left[ \mathbb{E} \left(\boldsymbol{\theta}_t \mid \boldsymbol{\theta}_{t+1}, \psi_t \right) \mid \boldsymbol{\psi}_T\right], && \\
    \boldsymbol{P}_{t \mid T} &= 
    \operatorname{Cov}( \boldsymbol{\theta}_{t} \mid \boldsymbol{\psi}_T ) 
    =
    \mathbb{E} \left[ \operatorname{Cov} \left( \boldsymbol{\theta}_{t} \mid \boldsymbol{\theta}_{t+1}, \boldsymbol{\psi}_t \right) \mid \boldsymbol{\psi}_T \right] + 
    \operatorname{Cov} \left[ \mathbb{E} \left( \boldsymbol{\theta}_{t} \mid \boldsymbol{\theta}_{t+1}, \boldsymbol{\psi}_t \right) \mid \boldsymbol{\psi}_T \right].
    &&
    \end{flalign*}

    \item
    To calculate $\mathbb{E} \left(\boldsymbol{\theta}_t \mid \boldsymbol{\theta}_{t+1}, \psi_t \right)$ and $\operatorname{Cov} \left(\boldsymbol{\theta}_t \mid \boldsymbol{\theta}_{t+1}, \psi_t \right)$ in above equations, consider the following approximated joint Gaussian distribution 
    \begin{align*}
            \left(\begin{array}{l}
            \boldsymbol{\theta}_t \\
            \boldsymbol{\theta}_{t+1}
            \end{array} \mid \boldsymbol{\psi}_{t} \right)
            \stackrel{\text{approx}}{\sim}  
            \operatorname{MN}\left(
            \begin{bmatrix}
            \boldsymbol{\theta}_{t \mid t} \\
           \boldsymbol{\theta}_{t+1 \mid t}
            \end{bmatrix},
            \begin{bmatrix}
            \boldsymbol{P}_{t \mid t}  & \boldsymbol{\Sigma}_{t , t+1}\\
           \boldsymbol{\Sigma}_{t , t+1}^T  & \boldsymbol{P}_{t+1 \mid t}
            \end{bmatrix}
            \right),
    \end{align*}    
    where $\boldsymbol{\theta}_{t \mid t}$ and $\boldsymbol{P}_{t \mid t}$ have been derived from MKF, and the rest parameters in the above multivariate Gaussian distribution can be calculated as follows:    
    \begin{flalign*}
        \boldsymbol{\theta}_{t+1 \mid t}  
        &= \mathbb{E} \left( \boldsymbol{\theta}_{t+1} \mid \boldsymbol{\psi}_{t} \right) && \\
        &= \sum_{p=0}^{1} \sum_{q=0}^{1} \mathbb{E} \left( \boldsymbol{\theta}_{t+1} \mid \boldsymbol{\psi}_{t}, I_t = p, I_{t+1} = q \right) \operatorname{Pr}(I_t = p, I_{t+1} = q \mid \boldsymbol{\psi}_{t}) &&\\
        &= \sum_{p=0}^{1} \sum_{q=0}^{1} \boldsymbol{\theta}_{t+1 \mid t}^{(p,q)} \operatorname{Pr}(I_t = p, I_{t+1} = q \mid \boldsymbol{\psi}_{t}), && \\
        \boldsymbol{\Sigma}_{t , t+1} 
        &= \operatorname{Cov} (\boldsymbol{\theta}_{t},\boldsymbol{\theta}_{t+1} \mid \boldsymbol{\psi}_{t}) &&\\ 
        &=
        \mathbb{E} (\boldsymbol{\theta}_{t} \boldsymbol{\theta}_{t+1}^T \mid \boldsymbol{\psi}_{t}) - \mathbb{E} ( \boldsymbol{\theta}_{t} \mid \boldsymbol{\psi}_{t} ) \mathbb{E} ( \boldsymbol{\theta}_{t+1} \mid \boldsymbol{\psi}_{t} )^T &&\\
        &=
        \mathbb{E} (\boldsymbol{\theta}_{t} \boldsymbol{\theta}_{t+1}^T \mid \boldsymbol{\psi}_{t}) - \boldsymbol{\theta}_{t \mid t} \boldsymbol{\theta}_{t+1 \mid t}^T, &&
    \end{flalign*}
    where the second term in the last equation above is calculated from MKF. Use the fact that $\boldsymbol{\theta}_{t}$ and $I_{t+1}$ are independent given $I_t$, $\boldsymbol{x}_i$ and $\hat{\boldsymbol{z}}_{1:n}^{\tau}$ at any iteration $\tau$. The first term in the above equation can be calculated as follows:
    \begin{flalign*}
        &\mathbb{E} (\boldsymbol{\theta}_{t} \boldsymbol{\theta}_{t+1}^T \mid \boldsymbol{\psi}_{t})
        =
        \mathbb{E} \left[ \mathbb{E} \left(\boldsymbol{\theta}_{t} \boldsymbol{\theta}_{t+1}^T \mid \boldsymbol{\psi}_{t}, I_t = p, I_{t+1}=q \right) \mid \boldsymbol{\psi}_{t} \right] && \\
        &=
        \mathbb{E} \left[ \mathbb{E} \left(\boldsymbol{\theta}_{t} \left( \boldsymbol{\gamma}_{I_{t+1},t+1} + \mathbf{G}_{I_{t+1},t+1} \boldsymbol{\theta}_{t} + \boldsymbol{w}_{I_{t+1},t+1} \right)^T \mid \boldsymbol{\psi}_{t}, I_t = p, I_{t+1}=q \right) \mid \boldsymbol{\psi}_{t} \right] && \\
        &=
        \mathbb{E} \left[ \mathbb{E} \left(\boldsymbol{\theta}_{t} \boldsymbol{\gamma}_{q,t+1}^T + \boldsymbol{\theta}_{t} \boldsymbol{\theta}_{t}^T \mathbf{G}_{q,t+1}^T  + \boldsymbol{\theta}_{t} \boldsymbol{w}_{q,t+1}^T  \mid \boldsymbol{\psi}_{t}, I_t = p, I_{t+1}=q \right) \mid \boldsymbol{\psi}_{t} \right] && \\
        &=
        \mathbb{E} \left[  \boldsymbol{\theta}_{t|t}^{(p)} \boldsymbol{\gamma}_{q,t+1}^T + \mathbb{E} \left( \boldsymbol{\theta}_{t} \boldsymbol{\theta}_{t}^T \mid \boldsymbol{\psi}_{t}, I_t = p, I_{t+1}=q \right) \mathbf{G}_{q,t+1}^T \mid \boldsymbol{\psi}_{t} \right] + 0 && \\
        &=
        \sum_{p=0}^{1} \sum_{q=0}^{1} \operatorname{Pr}(I_t = p, I_{t+1} = q \mid \boldsymbol{\psi}_{t}) \boldsymbol{\theta}_{t|t}^{(p)} \boldsymbol{\gamma}_{q,t+1}^T + \mathbb{E} \left[ \mathbb{E} \left( \boldsymbol{\theta}_{t} \boldsymbol{\theta}_{t}^T \mid \boldsymbol{\psi}_{t}, I_t = p, I_{t+1}=q \right) \mathbf{G}_{q,t+1}^T \mid \boldsymbol{\psi}_{t} \right] && \\
        &=
        \sum_{p=0}^{1} \sum_{q=0}^{1} \operatorname{Pr}(I_t = p, I_{t+1} = q \mid \boldsymbol{\psi}_{t}) \boldsymbol{\theta}_{t|t}^{(p)} \boldsymbol{\gamma}_{q,t+1}^T 
        + \mathbb{E} \left[ \left( \boldsymbol{\theta}_{t|t}^{(p)} {\boldsymbol{\theta}_{t|t}^{(p)}}^T + \boldsymbol{P}_{t|t}^{(p)} \right) \mathbf{G}_{q,t+1}^T \mid \boldsymbol{\psi}_{t} \right]
        && \\
        &=
        \sum_{p=0}^{1} \sum_{q=0}^{1} \operatorname{Pr}(I_t = p, I_{t+1} = q \mid \boldsymbol{\psi}_{t}) \boldsymbol{\theta}_{t|t}^{(p)} \boldsymbol{\gamma}_{q,t+1}^T &&\\
        &+ 
        \sum_{p=0}^{1} \sum_{q=0}^{1} \operatorname{Pr}(I_t = p, I_{t+1} = q \mid \boldsymbol{\psi}_{t}) \left[ \left( \boldsymbol{\theta}_{t|t}^{(p)} {\boldsymbol{\theta}_{t|t}^{(p)}}^T + \boldsymbol{P}_{t|t}^{(p)} \right) \mathbf{G}_{q,t+1}^T  \right]
        && \\
        &=
        \sum_{p=0}^{1} \sum_{q=0}^{1} \operatorname{Pr}(I_t = p, I_{t+1} = q \mid \boldsymbol{\psi}_{t}) \left[ \boldsymbol{\theta}_{t|t}^{(p)} \boldsymbol{\gamma}_{q,t+1}^T 
        + 
        \boldsymbol{\theta}_{t|t}^{(p)} {\boldsymbol{\theta}_{t|t}^{(p)}}^T \mathbf{G}_{q,t+1}^T + \boldsymbol{P}_{t|t}^{(p)}  \mathbf{G}_{q,t+1}^T  \right].
        &&
    \end{flalign*}

    The covariance matrix for $\boldsymbol{\theta}_{t+1}$ given $\boldsymbol{\psi}_t$ is calculated as follows:
    \begin{flalign*}
     &\boldsymbol{P}_{t+1 \mid t} = \operatorname{Cov} (\boldsymbol{\theta}_{t+1} \mid \boldsymbol{\psi}_{t}) && \\
    ={}&
    \mathbb{E} \left[ \operatorname{Cov} \left( \boldsymbol{\theta}_{t+1} \mid I_t = p, I_{t+1}=q, \boldsymbol{\psi}_t \right) \mid \boldsymbol{\psi}_t \right] + 
    \operatorname{Cov} \left[ \mathbb{E} \left( \boldsymbol{\theta}_{t+1} \mid I_t = p, I_{t+1}=q, \boldsymbol{\psi}_t \right) \mid \boldsymbol{\psi}_t \right] &&\\
    ={}&
    \mathbb{E} \left[ \operatorname{Cov} \left( \boldsymbol{\gamma}_{I_{t+1},t+1} + \mathbf{G}_{I_{t+1},t+1} \boldsymbol{\theta}_{t} + \boldsymbol{w}_{I_{t+1},t+1} \mid I_t = p, I_{t+1}=q, \boldsymbol{\psi}_t \right) \mid \boldsymbol{\psi}_t \right] \\
    +&
    \operatorname{Cov} \left[ \mathbb{E} \left( \boldsymbol{\gamma}_{I_{t+1},t+1} + \mathbf{G}_{I_{t+1},t+1} \boldsymbol{\theta}_{t} + \boldsymbol{w}_{I_{t+1},t+1} \mid I_t = p, I_{t+1}=q, \boldsymbol{\psi}_t \right) \mid \boldsymbol{\psi}_t \right] &&\\
    ={}&
    \mathbb{E} \left[ \operatorname{Cov} \left( \boldsymbol{\gamma}_{q,t+1} + \mathbf{G}_{q,t+1} \boldsymbol{\theta}_{t} + \boldsymbol{w}_{q,t+1} \mid I_t = p,  \boldsymbol{\psi}_t \right) \mid \boldsymbol{\psi}_t \right] \\
    +&
    \operatorname{Cov} \left[ \mathbb{E} \left( \boldsymbol{\gamma}_{q,t+1} + \mathbf{G}_{q,t+1} \boldsymbol{\theta}_{t} + \boldsymbol{w}_{q,t+1} \mid I_t = p,  \boldsymbol{\psi}_t \right) \mid \boldsymbol{\psi}_t \right] &&\\
    ={}&
    \mathbb{E} \left[  \mathbf{G}_{q,t+1} \boldsymbol{P}_{t|t}^{(p)} \mathbf{G}_{q,t+1}^T + \mathbf{W}_{q,t+1} \mid \boldsymbol{\psi}_t \right] + 
    \operatorname{Cov} \left[ \boldsymbol{\gamma}_{q,t+1} + \mathbf{G}_{q,t+1} \boldsymbol{\theta}_{t|t}^{(p)} \mid \boldsymbol{\psi}_t \right] &&\\
    ={}&
    \sum_{p=0}^{1} \sum_{q=0}^{1} \operatorname{Pr}(I_t = p, I_{t+1} = q \mid \boldsymbol{\psi}_{t}) \left[  \mathbf{G}_{q,t+1} \boldsymbol{P}_{t|t}^{(p)} \mathbf{G}_{q,t+1}^T + \mathbf{W}_{q,t+1} \right] \\
    +& 
    \operatorname{Cov} \left[ \boldsymbol{\gamma}_{q,t+1} + \mathbf{G}_{q,t+1} \boldsymbol{\theta}_{t|t}^{(p)} \mid \boldsymbol{\psi}_t \right],
    \end{flalign*}
    where the second term in the above equation is 
    \begin{flalign*}
    \operatorname{Cov} \left[ \boldsymbol{\gamma}_{q,t+1} + \mathbf{G}_{q,t+1} \boldsymbol{\theta}_{t|t}^{(p)} \mid \boldsymbol{\psi}_t \right] 
    &= 
    \mathbb{E} \left[ 
    \left( \boldsymbol{\gamma}_{q,t+1} + \mathbf{G}_{q,t+1} \boldsymbol{\theta}_{t|t}^{(p)} \right)
    \left( \boldsymbol{\gamma}_{q,t+1} + \mathbf{G}_{q,t+1} \boldsymbol{\theta}_{t|t}^{(p)} \right)^T
    \mid \boldsymbol{\psi}_t \right] \\
    &-
    \mathbb{E} \left( \boldsymbol{\gamma}_{q,t+1} + \mathbf{G}_{q,t+1} \boldsymbol{\theta}_{t|t}^{(p)} \mid \boldsymbol{\psi}_t \right)
    \mathbb{E} \left( \boldsymbol{\gamma}_{q,t+1} + \mathbf{G}_{q,t+1} \boldsymbol{\theta}_{t|t}^{(p)} \mid \boldsymbol{\psi}_t \right)^T.
    && \\
    \end{flalign*}
    For the first term,
    \begin{flalign*}
        &\mathbb{E} \left[ 
        \left( \boldsymbol{\gamma}_{q,t+1} + \mathbf{G}_{q,t+1} \boldsymbol{\theta}_{t|t}^{(p)} \right)
        \left( \boldsymbol{\gamma}_{q,t+1} + \mathbf{G}_{q,t+1} \boldsymbol{\theta}_{t|t}^{(p)} \right)^T
        \mid \boldsymbol{\psi}_t \right] 
         && \\
        =&\mathbb{E} \left[
        \boldsymbol{\gamma}_{q,t+1} \boldsymbol{\gamma}_{q,t+1}^T + \boldsymbol{\gamma}_{q,t+1} {\boldsymbol{\theta}_{t|t}^{(p)}}^T \mathbf{G}_{q,t+1}^T + \mathbf{G}_{q,t+1} \boldsymbol{\theta}_{t|t}^{(p)} \boldsymbol{\gamma}_{q,t+1}^T + \mathbf{G}_{q,t+1} \boldsymbol{\theta}_{t|t}^{(p)} {\boldsymbol{\theta}_{t|t}^{(p)}}^T \mathbf{G}_{q,t+1}^T
        \mid \boldsymbol{\psi}_t \right] && \\
        = &
        \sum_{p=0}^{1} \sum_{q=0}^{1} \operatorname{Pr}(I_t = p, I_{t+1} = q \mid \boldsymbol{\psi}_{t}) \left[ \boldsymbol{\gamma}_{q,t+1} \boldsymbol{\gamma}_{q,t+1}^T + \boldsymbol{\gamma}_{q,t+1} {\boldsymbol{\theta}_{t|t}^{(p)}}^T \mathbf{G}_{q,t+1}^T \right. &&\\
        & \left.+ \mathbf{G}_{q,t+1} \boldsymbol{\theta}_{t|t}^{(p)} \boldsymbol{\gamma}_{q,t+1}^T + \mathbf{G}_{q,t+1} \boldsymbol{\theta}_{t|t}^{(p)} {\boldsymbol{\theta}_{t|t}^{(p)}}^T \mathbf{G}_{q,t+1}^T \right].
    \end{flalign*}
    
    For the second term,
    \begin{flalign*}
        \mathbb{E} \left( \boldsymbol{\gamma}_{q,t+1} + \mathbf{G}_{q,t+1} \boldsymbol{\theta}_{t|t}^{(p)} \mid \boldsymbol{\psi}_t \right) 
        =
        \sum_{p=0}^{1} \sum_{q=0}^{1} \operatorname{Pr}(I_t = p, I_{t+1} = q \mid \boldsymbol{\psi}_{t}) \left[ \boldsymbol{\gamma}_{q,t+1} + \mathbf{G}_{q,t+1} \boldsymbol{\theta}_{t|t}^{(p)} \right].
    \end{flalign*}

\item
Now we have all elements in the joint distribution, the conditional expectation and covariance can be calculated as
\begin{flalign*}
    \mathbb{E} \left(\boldsymbol{\theta}_t \mid \boldsymbol{\theta}_{t+1}, \psi_t \right)
    &=
    \boldsymbol{\theta}_{t|t} + \boldsymbol{\Sigma}_{t,t+1} \boldsymbol{P}_{t+1|t}^{-1} (\boldsymbol{\theta}_{t+1}-\boldsymbol{\theta}_{t+1|t}), && \\
    \operatorname{Cov} \left(\boldsymbol{\theta}_t \mid \boldsymbol{\theta}_{t+1}, \psi_t \right)
    &=
    \boldsymbol{P}_{t|t} - \boldsymbol{\Sigma}_{t,t+1} \boldsymbol{P}_{t+1|t}^{-1} \boldsymbol{\Sigma}_{t,t+1}^T.
\end{flalign*}

\item
Now we can calculate the marginal distribution
\begin{flalign*}
\boldsymbol{\theta}_{t \mid T} &=
\mathbb{E}\left[ \mathbb{E} \left(\boldsymbol{\theta}_t \mid \boldsymbol{\theta}_{t+1}, \psi_t \right) \mid \boldsymbol{\psi}_T\right] && \\
&=
\boldsymbol{\theta}_{t|t} + \boldsymbol{\Sigma}_{t,t+1} \boldsymbol{P}_{t+1|t}^{-1} (\mathbb{E}\left[ \boldsymbol{\theta}_{t+1} \mid \boldsymbol{\psi}_T\right] -\boldsymbol{\theta}_{t+1|t}) &&\\
&=
\boldsymbol{\theta}_{t|t} + \boldsymbol{\Sigma}_{t,t+1} \boldsymbol{P}_{t+1|t}^{-1} ( \boldsymbol{\theta}_{t+1|T}  -\boldsymbol{\theta}_{t+1|t}), \\
\boldsymbol{P}_{t \mid T} 
&= 
\mathbb{E} \left[ \operatorname{Cov} \left( \boldsymbol{\theta}_{t} \mid \boldsymbol{\theta}_{t+1}, \boldsymbol{\psi}_t \right) \mid \boldsymbol{\psi}_T \right] + 
\operatorname{Cov} \left[ \mathbb{E} \left( \boldsymbol{\theta}_{t} \mid \boldsymbol{\theta}_{t+1}, \boldsymbol{\psi}_t \right) \mid \boldsymbol{\psi}_T \right] && \\
&=
\mathbb{E} \left[ \boldsymbol{P}_{t|t} - \boldsymbol{\Sigma}_{t,t+1} \boldsymbol{P}_{t+1|t}^{-1} \boldsymbol{\Sigma}_{t,t+1}^T \mid \boldsymbol{\psi}_T \right] + 
\operatorname{Cov} \left[ \boldsymbol{\theta}_{t|t} + \boldsymbol{\Sigma}_{t,t+1} \boldsymbol{P}_{t+1|t}^{-1} (\boldsymbol{\theta}_{t+1}-\boldsymbol{\theta}_{t+1|t}) \mid \boldsymbol{\psi}_T \right] && \\
&=
\boldsymbol{P}_{t|t} - \boldsymbol{\Sigma}_{t,t+1} \boldsymbol{P}_{t+1|t}^{-1} \boldsymbol{\Sigma}_{t,t+1}^T + 
\boldsymbol{\Sigma}_{t,t+1} \boldsymbol{P}_{t+1|t}^{-1} \operatorname{Cov}(\boldsymbol{\theta}_{t+1} \mid \boldsymbol{\psi}_T)
\boldsymbol{P}_{t+1|t}^{-1} \boldsymbol{\Sigma}_{t,t+1}^T &&\\
&=
\boldsymbol{P}_{t|t} - \boldsymbol{\Sigma}_{t,t+1} \boldsymbol{P}_{t+1|t}^{-1} \boldsymbol{\Sigma}_{t,t+1}^T + 
\boldsymbol{\Sigma}_{t,t+1} \boldsymbol{P}_{t+1|t}^{-1} \boldsymbol{P}_{t+1|T}
\boldsymbol{P}_{t+1|t}^{-1} \boldsymbol{\Sigma}_{t,t+1}^T. &&
\end{flalign*}

\item We compute the probability of interest $\operatorname{Pr}(I_{t}=p \mid  \boldsymbol{\psi}_{T})$ as follows
\begin{enumerate}
    \item \label{MFIS1_proof}
    \begin{flalign*}
        \operatorname{Pr}(I_{t}=p, I_{t+1}=q \mid  \boldsymbol{\psi}_{T}) 
        &= \operatorname{Pr}(I_{t+1}=q \mid  \boldsymbol{\psi}_{T}) \operatorname{Pr}(I_{t}=p \mid I_{t+1}=q,   \boldsymbol{\psi}_{T}) &&\\
        & \approx \operatorname{Pr}(I_{t+1}=q \mid  \boldsymbol{\psi}_{T}) \operatorname{Pr}(I_{t}=p \mid I_{t+1}=q,   \boldsymbol{\psi}_{t}) &&\\
        &= \frac{\operatorname{Pr}(I_{t+1}=q \mid  \boldsymbol{\psi}_{T}) \operatorname{Pr}(I_{t}=p \mid   \boldsymbol{\psi}_{t}) \operatorname{Pr}(I_{t+1}=q \mid I_{t}=p,   \boldsymbol{\psi}_{t})}{\operatorname{Pr}(I_{t+1}=q \mid \boldsymbol{\psi}_{t})},  &&
    \end{flalign*}
    where the denominator can be obtained by summing up the probability in MKF (\ref{MKFprob1_proof}):
    $$
    \sum_{p=0}^{1} \operatorname{Pr}(I_{t}=p, I_{t+1}=q \mid \boldsymbol{\psi}_{t}) = \operatorname{Pr}(I_{t+1}=q \mid \boldsymbol{\psi}_{t}).
    $$
    
    The second step above involves an approximation which becomes exact if and only if $\operatorname{Pr}(\boldsymbol{\psi}_{t+1:T} \mid  I_{t}=p, I_{t+1}=q ,\boldsymbol{\psi}_{t} ) $ = $\operatorname{Pr}(\boldsymbol{\psi}_{t+1:T} \mid  I_{t+1}=q ,\boldsymbol{\psi}_{t} ) $. To see this, expand the second term on the right-hand side of the first equation as
    \begin{flalign*}
        \operatorname{Pr}(I_{t}=p \mid I_{t+1}=q,   \boldsymbol{\psi}_{T}) 
        &= \operatorname{Pr}(I_{t}=p \mid I_{t+1}=q,   \boldsymbol{\psi}_{t}, \boldsymbol{\psi}_{t+1:T}) && \\
        &= \frac{ \operatorname{Pr}(I_{t}=p , \boldsymbol{\psi}_{t+1:T} \mid I_{t+1}=q,   \boldsymbol{\psi}_{t}) }{  \operatorname{Pr}(\boldsymbol{\psi}_{t+1:T} \mid I_{t+1}=q,   \boldsymbol{\psi}_{t})   } && \\
        &= \frac{ \operatorname{Pr}(I_{t}=p \mid I_{t+1}=q,   \boldsymbol{\psi}_{t}) \operatorname{Pr}(\boldsymbol{\psi}_{t+1:T} \mid I_{t}=p, I_{t+1}=q,   \boldsymbol{\psi}_{t}) }{ \operatorname{Pr}(\boldsymbol{\psi}_{t+1:T} \mid I_{t+1}=q,   \boldsymbol{\psi}_{t}) }.
    \end{flalign*}

    We can see that the approximation becomes equality only if the above condition holds.
    
    \item 
    $
    \operatorname{Pr}(I_{t}=p \mid  \boldsymbol{\psi}_{T}) = \sum_{q=0}^{1} \operatorname{Pr}(I_{t}=p,I_{t+1}=q \mid \boldsymbol{\psi}_{T})
    $.    

\end{enumerate}

\end{enumerate}

\end{proof}

\bibliographystyle{unsrtnat} \bibliography{bibli}

\end{document}